\documentclass[prl,twocolumn,showpacs,preprintnumbers,amsmath,amssymb,footinbib]{revtex4-1}

\usepackage{graphicx}
\usepackage[colorlinks=true]{hyperref}
\usepackage{psfrag}
\usepackage{dcolumn}
\usepackage{bm}
\usepackage{color}
\usepackage{array}
\usepackage{longtable}
\usepackage{dcolumn}

\def\ket#1{\mathinner{|{#1}\rangle}}
\def\braket#1{\mathinner{\langle{#1}\rangle}}

\begin{document}
\date{\today}
\author{Simon E. Nigg, Hanhee Paik, Brian
  Vlastakis, Gerhard Kirchmair, Shyam~Shankar, Luigi~Frunzio, Michel Devoret,
  Robert Schoelkopf and Steven Girvin}
\affiliation{Departments of Physics and Applied Physics, Yale University, New Haven, CT
  06520, USA}
\begin{abstract}
We present a semi-classical method for determining the effective low-energy quantum Hamiltonian
of weakly anharmonic superconducting circuits containing mesoscopic Josephson
junctions coupled to electromagnetic environments made of an arbitrary combination of distributed
and lumped elements. A convenient basis, capturing the multi-mode physics, is given by
the quantized eigenmodes of the linearized circuit and is fully determined by a classical linear response function. The method is
used to calculate numerically the low-energy spectrum of a 3D-transmon
system, and quantitative agreement with measurements is found.
\end{abstract}
\title{Black-box superconducting circuit quantization}
\maketitle

Superconducting electronic circuits containing nonlinear elements such as Josephson
junctions (JJs) are of interest for quantum information
processing~\cite{Devoret-2004a,Wallraff-2004a}, due to their
nonlinearity and weak intrinsic dissipation. The discrete
low-energy spectrum of such circuits can now be measured to a
precision of better than one part per million~\cite{Paik-2011a}. The question
thus naturally arises of how well one can theoretically model such man-made artificial
atoms. Indeed, increasing evidence indicates that due to increased
coupling strengths~\cite{Devoret-2007a}, current models are
reaching their
limits~\cite{Houck-2008a,Bourassa-2009a,Niemczyk-2010a,Filipp-2011a,Viehmann-2011a} and in
order to further our ability to design, optimize and manipulate these systems, developing models beyond these limits becomes
necessary. This is the goal of the
present work.

An isolated ideal JJ has only one collective degree of freedom:
the order parameter phase difference $\varphi$ across the
junction. The zero-temperature, sub-gap physics of this system, with
Josephson energy $E_J$ and charging energy $E_C$, is described by the
Cooper-pair box Hamiltonian 
\begin{equation}\label{eq:8}
H_{\rm CPB} = 4E_C(\hat N-N_g)^2-E_J\cos(\hat\varphi),
\end{equation}
where
$\hat N$ is the Cooper-pair number operator conjugate to
$\hat\varphi$ and $N_g$ an offset charge. This model is exactly solvable in terms of Mathieu
functions~\cite{ACottetThesis,Koch-2007a}. The crucial
feature that emerges from this solution is that the charge dispersion,
i.e. the maximal variation of the eigenenergies with $N_g$, is {\em exponentially} suppressed with $E_J/E_C$ while the
relative anharmonicity decreases only algebraically with a slow power-law in
$E_J/E_C$. As a consequence, there exists a regime with $E_J\gg E_C$ --
the transmon regime -- where the anharmonicity is much larger than the
linewidth (e.g. due to fluctuation of the offset charge $N_g$), thus satisfying the operability
condition of a qubit~\cite{Schreier-2008a}. This is the regime of interest
here.

In order to be useful for quantum information processing tasks,
several Josephson qubits must be made to controllably interact with each other and
spurious interactions with uncontrolled (environmental) degrees of
freedom must be minimized. In circuit quantum
electrodynamics~\cite{Blais-2004a,Wallraff-2004a,Koch-2007a} (cQED),
this is achieved by coupling the JJs to a common microwave environment with a desired
discrete mode structure. So far such systems have mostly been described theoretically by models well
known from quantum optics such as the single-mode Jaynes-Cummings model and
extensions thereof~\cite{Jaynes-1963a,*Tavis-1968a}.

When applied to
\begin{figure}
\begin{center}
\includegraphics[width=0.9\columnwidth]{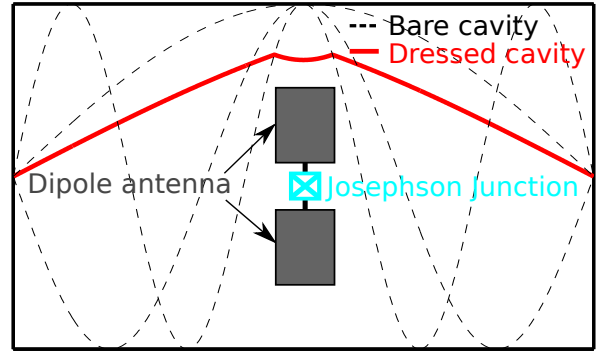}\caption{(Color online)
  Cartoon of a JJ at the center of a broadband
  dipole antenna inside a 3D microwave cavity. The presence of
  the antenna alters the geometry of the cavity-mode (full (red)
  curve) and a precise description requires the
inclusion of many bare modes (dashed curves).\label{fig:2}}
\end{center}
\end{figure}
superconducting circuits with multi-level artificial atoms,
multi-mode cavities and increased coupling
strengths~\cite{Devoret-2007a,Bourassa-2009a,Niemczyk-2010a} however, several
technical and practical difficulties with these approaches arise. For
example, capturing important effects of non-computational qubit
states requires going to high orders in perturbation theory~\cite{DiCarlo-2009a}. Also, determining
the {\em bare} Hamiltonian parameters, in terms of which these models
are defined, is cumbersome and requires iterating
between experiment and theory. Perhaps even more important are the
shortcomings of the traditional approaches in dealing with the
multiple modes of the cavity. Indeed high-energy, off-resonant cavity modes have already been measured to contribute
substantially to the inter-qubit
interaction strength~\cite{DiCarlo-2009a,Filipp-2011a} and, via
the multi-mode Purcell effect, also to affect the coherence properties (relaxation and dephasing) of the
qubits~\cite{Houck-2008a}. Attempts at including this multi-mode physics in the standard
models however, lead to difficulties with diverging
series and QED renormalization issues~\cite{Filipp-2011a,Bourassa-2011-unpub}, which to
the best of our knowledge remain unresolved. Fig.~\ref{fig:2} illustrates the origin of
the problem with the example of a JJ inside a 3D
cavity (3D-transmon)~\cite{Paik-2011a}. The presence of a relatively large
metallic dipole antenna~\footnote{In current realizations of the 3D-transmon
  qubits, the length of the antenna is between $1$ and $10\%$ of the
  wavelength of the fundamental bare cavity mode.} can
strongly alter the geometry of the cavity modes. This essentially
classical effect, can be
accounted for precisely only by including a sufficiently large number
of bare modes.

In contrast, we propose to start by considering the coupled but
{\em linearized} problem in order to find a
basis that incorporates the main effects of the coupling between
multi-level qubits and a multi-mode cavity and then
account for the weak anharmonicity of the Josephson potential perturbatively. The crucial
assumption made here 
is that
charge dispersion effects can be safely neglected. This is
reasonable given that in
state-of-the art implementations of transmon
qubits~\cite{Paik-2011a,Reed-2012a}, charge dispersion only
contributes a negligible amount to the measured linewidths. Previous
work discussed the nonlinear dynamics of a
JJ embedded in an external circuit classically~\cite{Manucharyan-2007a}. Here
we go one step further and show how the  knowledge of a classical, in
principle measurable,
linear response function lets us quantize the circuit, treating qubits
and cavity on equal footing.

\begin{figure}[ht]
\begin{center}
\def\svgwidth{0.32\columnwidth}
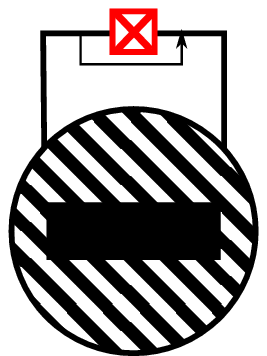\hfill\def\svgwidth{0.32\columnwidth}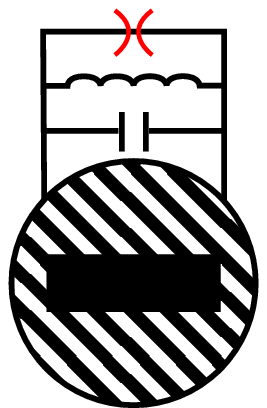\hfill\def\svgwidth{0.32\columnwidth}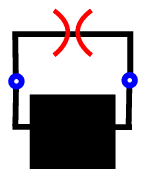\\\vspace*{0.5cm}
\def\svgwidth{\columnwidth}
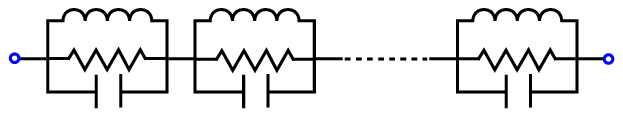\caption{(Color
  online) (a) Schematics
of a JJ ((red) boxed cross) coupled to an
arbitrary linear circuit (striped disk). (b) The Josephson element is
replaced by a parallel combination of: a linear inductance $L_J$, a
linear capacitance $C_J$ and a purely nonlinear element with energy
$E_J(1-\cos(\varphi))-(E_J/2)\varphi^2$,
represented by
the spider symbol. (c) The linear part of the circuit shown in (b) is
lumped into an impedance $Z(\omega)$ seen by the nonlinear element. (d)
Foster-equivalent circuit (pole-decomposition) of the impedance $Z(\omega)$.\label{fig:1}}
\end{center}
\end{figure}

{\it Single junction case.} We consider a system with a JJ with {\em bare} Josephson energy $E_J$ and charging energy $E_C$,
in parallel with a linear but otherwise arbitrary electromagnetic
environment as depicted in Fig.~\ref{fig:1} (a). Neglecting
dissipation, the unbiased junction alone is described
by the Hamiltonian~(\ref{eq:8}). At low energies, when $E_J\gg E_C$, quantum fluctuations
of the phase $\varphi$ across the junction are small compared with
$\pi$ and, as emphasized in the introduction, the probability of quantum
tunneling of the phase between minima of the cosine potential is negligibly
small. It is then reasonable to expand the latter in powers of $\varphi$,
thus obtaining the approximate circuit representation of Fig.~\ref{fig:1}~(b), in
which the spider symbol~\cite{Manucharyan-2007a} represents the purely nonlinear part and
$L_J={\phi_0}^2/E_J$ and $C_J=e^2/(2E_C)$ the
linear parts of the Josephson
element. Here $\phi_0=\hbar/(2e)$ is the reduced flux quantum. To
leading order, the energy of the spider element is given by
$E_{\rm nl}=-{\phi_0}^2\varphi^4/(24L_J)$. 

A quantity of central importance in the following is the impedance $Z(\omega)$ of the linear part of the
circuit depicted in Fig.~\ref{fig:1}~(c). The latter is a complex meromorphic function
and by virtue of Foster's theorem~\cite{Foster-1924,PMC-1948} can be
synthesized by the equivalent circuit of  parallel LCR oscillators in series shown in
Fig.~\ref{fig:1}~(d). Explicitly
\begin{equation}\label{eq:4}
Z(\omega) = \sum_{p=1}^M\left(j\omega
    C_p+\frac{1}{j\omega L_p}+\frac{1}{R_p}\right)^{-1},
\end{equation}
where $M$ is the number of modes~\footnote{The case of infinitely many discrete modes
  necessitates an extension of Foster's theorem as discussed
  in~\cite{Zinn-1951a}, but the results presented here still apply.} and we have adopted the electrical engineering
convention of writing the imaginary unit as $j=-i$. This equivalent circuit mapping corresponds, in electrical
engineering language, to diagonalizing the
linearized system of coupled harmonic oscillators. The resonance
frequencies of the linear circuit are determined by the real parts of
the poles of $Z$
or more conveniently by the real parts of the zeros of the admittance defined as
$Y(\omega)=Z(\omega)^{-1}$, and for weak dissipation,
i.e. $R_p\gg\sqrt{L_p/C_p}$, are given by
$\omega_p=(L_pC_p)^{-\frac{1}{2}}$. The imaginary parts of
the roots $(2R_pC_p)^{-1}$, give the resonances a finite width. The
effective resistances are given by $R_p=1/{\rm Re}Y(\omega_p)$ and the effective capacitances are
determined by the frequency derivative on resonance of the admittance as $C_p =(1/2)
{\rm Im}Y'(\omega_p)$. Here and in the following the prime stands for
the derivative with respect to frequency. Note that ${\rm
  Im}Y'(\omega) >0$~\cite{Foster-1924}. Together this yields a
compact expression for the quality
factor of mode $p$:
\begin{equation}\label{eq:5}
Q_p = \frac{\omega_p}{2}\frac{{\rm Im}Y'(\omega_p)}{{\rm Re}Y(\omega_p)}.
\end{equation}
When applied to the mode representing the qubit, Eq.~(\ref{eq:5}) gives an estimate
for the Purcell limit on the qubit lifetime $T_1=Q_{\rm
  qb}/\omega_{\rm qb}$
due to photons leaking out of the cavity.

In order to derive the effective low-energy quantum Hamiltonian of the
circuit, we
next neglect dissipation ($R_p\rightarrow\infty$) and introduce the normal (flux) coordinates
$\phi_p(t)=f_pe^{j\omega_p
  t}+(f_p)^*e^{-j\omega_pt}$ associated with each
LC oscillator in the equivalent circuit. We can then
immediately write the classical Hamiltonian function of
the equivalent circuit as
$\mathcal{H}_0=2\sum_{p=1}^M(f_p)^*(L_p)^{-1}f_p$,
where the
subscript $0$ indicates that we consider the linear part of the circuit. Kirchhoff's voltage law
implies that up to an
arbitrary constant,
$\phi(t)=\sum_{p=1}^M\phi_p(t)$, where $\phi(t)=\int_{-\infty}^tV(\tau)d\tau$
is the flux coordinate of the junction with voltage $V(t)$. Note that
by the second Josephson relation, the order parameter phase difference
is related to the latter via $\varphi(t)=\phi(t)/\phi_0$ (modulo $2\pi$).

Quantization is achieved in the canonical way~\cite{Devoret-1995a,Clerk-2010a} by replacing the
flux amplitudes of the equivalent oscillators by operators as
\begin{equation}\label{eq:6}
  f_p^{(*)}\rightarrow\sqrt{\frac{\hbar}{2}\mathcal{Z}_p^{\rm
      eff}}\,a_p^{(\dagger)},\quad \mathcal{Z}_{p}^{\rm
    eff}
  =\frac{2}{\omega_p{\rm
        Im}Y'(\omega_p)},
\end{equation}
with the dimensionless bosonic annihilation (creation) operators
$a_p$ ($a_p^{\dagger}$). Direct substitution yields the Hamiltonian
$H_0=\sum_{p}\hbar\omega_pa_p^{\dagger}a_p$ of $M$ uncoupled harmonic
oscillators (omitting the zero-point energies) and the Schr{\"o}dinger
operator of flux across the junction is
\begin{equation}\label{eq:7}
\hat\phi = \sum_{p=1}^M\sqrt{\frac{\hbar}{2}\mathcal{Z}_{p}^{{\rm
      eff}}}\left(a_p+a_p^{\dagger}\right).
\end{equation}
We emphasize that the harmonic modes $a_p$
represent collective excitations of the linear circuit and
their frequencies $\omega_p$ are the equivalent of dressed oscillator
frequencies. The coupling in the linear circuit is treated exactly and
in particular no rotating wave approximation is used.
\begin{table*}[ht]
\begin{ruledtabular}
\begin{tabular}{d@{}dd@{}dd@{}dr@{}ld@{}d|dd}
\multicolumn{2}{c}{$\nu_{01}$ (GHz)}&\multicolumn{2}{c}{$\nu_c$ (GHz)}&\multicolumn{2}{c}{$\nu_{02}$ 
(GHz)}&\multicolumn{2}{c}{$\alpha_{\rm qb}$
(MHz)}&\multicolumn{2}{c}{$\chi$ (MHz)}&\multicolumn{1}{c}{$L_J$
(nH)} &\multicolumn{1}{c}{$C_J$ (ff)}\\
\hline 7.77&(7.763) & 8.102&(8.105) & 15.33&(15.333) & -210&(-193) &
-90&(-80.6)& 5.83 & 7.6\\
7.544&(7.54) & 8.126&(8.05) & 14.808&(14.830) &-280&(-249)
&-30&(-33.0)& 6.12 & 9.2\\
7.376&(7.376) &7.858&(7.864) & 14.489&(14.495) &-264&(-257) &
-37.5&(-38.7)&6.67 & 4.0\\
7.058&(7.045) & 8.005&(8.023) & 13.788&(13.794) & -328&(-295) &
-13.2&(-13.3)& 7.45& 5.2\\
6.808&(6.793) & 8.019&(8.017) & 13.286&(13.294) & -330&(-293) & -8&(-8.4)&7.71
& 7.8\\
6.384&(6.386) & 7.832&(7.823) & 12.45&(12.449) & -318&(-324) &
-5.4&(-7.6)&9.40 & 0.34\\
\end{tabular}
\end{ruledtabular}\caption{Low-energy spectrum ($\nu_{01}$,
  $\nu_c$, $\nu_{02}$), qubit anharmonicity ($\alpha_{\rm qb}$) and
  state-dependent cavity shift ($\chi$) of six 3D-transmons. Results are shown in the
  format: experiment (theory). The theory values are obtained from a least
  square fit in $C_J$ of the numerically computed lowest three energy levels of the
  $\phi^6$ model. The fitted values of $C_J$ are given in the last
column. Their order of magnitude (a few femto-farads) agrees with estimates based
on the sizes of the junctions. The Josephson inductances $L_J$ are obtained from
  room-temperature resistance measurements of the junctions. \label{tab:1}}
\end{table*}

The Hamiltonian of the circuit including the JJ is
then $H=H_0+H_{\rm nl}$, where $H_{\rm
  nl}=-(\hat\phi)^4/(24{\phi_0}^2L_J)+\mathcal{O}((\hat\phi/\phi_0)^6)$. Physical
insight may be gained by treating the nonlinear terms as a
perturbation on top of $H_0$ assuming the eigenstates
$\ket{n_1,n_2,\dots,n_M}$ of the latter with energies
$E_{n_1,n_2,\dots,n_M}^{(0)}=\sum_in_i\hbar\omega_i$, to be non-degenerate. Considering only
the leading order $\phi^4$ nonlinearity, one then obtains the
reduced Hamiltonian
\begin{equation}
H_{4}=H_0'+\frac{1}{2}\sum_{pp'}\chi_{pp'}\hat n_p\hat n_{p'}.
\end{equation}
Here $\hat n_p=a_p^{\dagger}a_p$ and $H_0' = H_0+\sum_p\Delta_p\hat n_p$ includes a correction to the
Lamb-shift given by $\Delta_p=-\frac{e^2}{2L_J}\left(\mathcal{Z}^{\rm
    eff}_p\sum_q\mathcal{Z}^{\rm eff}_q-(\mathcal{Z}^{\rm
    eff}_p)^2/2\right)$. We have further introduced the generalized
$\chi$-shift $\chi_{pp'}$ between
modes $p$ and $p'$. Clearly, $\alpha_p\equiv\chi_{pp}$ is the anharmonicity
of the first excited state (self-Kerr)
of mode $p$ while $\chi_{pp'}=\chi_{p'p}$ with $p\not=p'$ is the state-dependent
frequency shift per excitation (cross-Kerr) of mode $p$ due to the presence of a single excitation
in mode $p'$. Explicitly we find
\begin{equation}\label{eq:1}
\chi_{pp}=-\frac{L_p}{L_J}\frac{C_J}{C_p}E_C,\quad\chi_{pp'}=-2\sqrt{\chi_{pp}\chi_{p'p'}}.
\end{equation}
Note that all modes acquire some anharmonicity due to the presence of
the nonlinear JJ. There is thus no strict separation of qubit and
cavity anymore. Colloquially, a mode with strong (weak) anharmonicity
will be called qubit-like (cavity-like). Interestingly, in this lowest
order approximation, the anharmonicity of mode $p$ is seen to be
proportional to the inductive participation
ratios~\cite{Manucharyan-2007a} $i_p\equiv L_p/L_J$ and inversely proportional to the
capacitive participation ratio $c_p\equiv C_p/C_J$. In the absence of
a galvanic short of the junction in the resonator circuit, as is the
case e.g. for a transmon qubit capacitively coupled to a cavity, it follows from the sum
rule $\lim_{\omega\rightarrow
  0}\left[Z(\omega)/(j\omega)\right]=\sum_pL_p = L_J$ that $i_p\leq
1$. Similarly, in the absence of any capacitance in series with $C_J$, it follows that $c_p\geq 1$, because $\lim_{\omega\rightarrow 0}\left[j\omega Z(\omega)\right]=\sum_pC_p^{-1}=C_{\Sigma}^{-1}$,
where $C_{\Sigma}=C_J+C_{\parallel}$  and $C_{\parallel}$ is the total
capacitance in parallel with $C_J$. Hence
we see that in this experimentally relevant case, the effective
anharmonicity of the qubit-like mode is
always reduced as compared with the anharmonicity of the bare qubit
given by $-E_C$~\cite{Koch-2007a}. Remarkably, in this approximation
we find (see Eq.~(\ref{eq:1})) that the cross-Kerr shift
between two modes is twice the geometric mean of the anharmonicities of
the two modes. 

We emphasize that
the above expressions do not however account for higher order effects in
anharmonicity such as the change of sign of the cross-Kerr shift observed
in the straddling regime~\cite{Koch-2007a,Boissonneault-2010a}. Such
effects are however fully captured by the full
model $H=H_0+H_{\rm nl}$, which can be solved numerically. Remarkably,
because the {\em dressed} modes already resum all the {\em bare}
harmonic modes, typically only a few dressed modes $M^*\ll M$ need to be
included for good convergence, thus considerably reducing the size of the effective Hilbert
space, which scales as $\prod_{p=1}^{M^*}(N_p+1)$ where $N_p$ is the maximal allowed number
of excitations in mode $p$ (e.g. $N_p=1$ in a two-level approximation).

{\em Charge dispersion.} 
 By assumption charge dispersion effects are
 neglected in the above approach. One may however ask how the charge
 dispersion of an {\em isolated} JJ is affected when the latter is
 coupled to a cavity. As in the Caldeira-Leggett model~\cite{Caldeira-1981a}, the
 coupling between the JJ and Harmonic oscillators suppresses the
 probability of flux tunneling and hence reduces charge
 dispersion of the qubit further. A simple estimate of the suppression
 factor is provided by the
 probability $P_0$ of leaving the circuit in the ground state after a flux
 tunneling event and is found to be given
 by the ``Lamb-M{\"o}ssbauer'' factor $P_0
\approx e^{-\frac{1}{2}\sum_{p\not={\rm qb}}\left(\frac{{\delta
         q}^2}{2C_p}\right)\big/(\hbar\omega_p)}$, where the sum
 excludes the qubit mode and $\delta
 q=C_J\phi_0/\tau$ is the charge (momentum) kick generated by a
 $\phi_0$ flux slip through the JJ of duration $\tau$ and $C_p =
 (1/2){\rm Im} Y'(\omega_p)$. Thus our
 assumption of neglecting charge dispersion of the qubit is well justified.

Interestingly though, each eigenmode
 of the system inherits some charge dispersion. This effect,
 essentially a consequence of hybridization, is of
 particular importance for applications such as quantum information
 storage in high-Q cavities coupled to
 JJs and is the subject of
work in progress.

{\it Generalization to $N$ junctions.} The approach can be extended to
circuits with multiple JJs connected in parallel to a common linear circuit. Details
about the derivation are given in the supplementary
material~\cite{supp-mat-BBQ} and we here only state the results. For
$N$ qubits, the resonance frequencies of the linear part of the
circuit are determined by the
zeros of the admittance $Y_k(\omega)\equiv Z_{kk}(\omega)^{-1}$
for any choice of reference port $k=1,\dots ,N$,
where $\mathbf{Z}$ is the $N\times N$ impedance matrix of the linear part
of the circuit with a port being associated with each junction. The
flux operators of the $N$ junctions, with reference port $k$, are
given by ($l=1,\dots,N$)
\begin{equation}\label{eq:7}
\hat\phi_l^{(k)} =
\sum_{p=1}^M\frac{Z_{lk}(\omega_p)}{Z_{kk}(\omega_p)}\sqrt{\frac{\hbar}{2}\mathcal{Z}_{kp}^{{\rm
      eff}}}\left(a_p+a_p^{\dagger}\right),
\end{equation}
where $\mathcal{Z}_{kp}^{\rm eff}=2/[\omega_p{\rm
  Im}Y'_{k}(\omega_p)]$. Note that the resonance frequencies are independent of the choice of
reference port, while the eigenmodes do depend on it. In lowest
order of PT and in the $\phi^4$ approximation, we find
\begin{align}\label{eq:3}
\alpha_{p}&=-12\beta_{pppp},\quad
\chi_{qp}=-24\beta_{qqpp},\quad q\not=p,
\end{align}
as well as the correction to the Lamb-shift $\Delta_p=6\beta_{pppp}-12\sum_{q}\beta_{qqpp}$. Here
$\beta_{qq'pp'}=\sum_{s=1}^N\frac{e^2}{24L_J^{(s)}}\xi_{sq}\xi_{sq'}\xi_{sp}\xi_{sp'}$,
and choosing the first port as the reference port ($k=1$), $\xi_{sp}=\frac{Z_{s1}(\omega_p)}{Z_{11}(\omega_p)}\sqrt{\mathcal{Z}_{1p}^{{\rm
      eff}}}$. Notice that the Cauchy-Schwarz inequality implies that
$|\chi_{qp}|\leq 2\sqrt{\alpha_{q}\alpha_{p}}$. Also, if $q$ and $q'$ refer to two
different qubit-like modes, then $\chi_{qq'}$ is a measure for the {\em
  total} interaction strength (cavity mediated and direct dipole-dipole coupling) between these two qubits. 


{\em Comparison with experiment.} As a demonstration of this method, we apply it to
the case illustrated in Fig.~\ref{fig:2} of a single JJ coupled to a 3D cavity~\cite{Paik-2011a}. The admittance at the junction port $Y$ is a parallel combination of the
linearized qubit admittance and the admittance $Y_c$ of the
cavity-antenna system, i.e. $Y(\omega)=j\omega C_J-j/(\omega
L_J)+Y_c(\omega)$. The junction is
assumed to be dissipationless corresponding to a Purcell-limited
qubit and ohmic losses of the cavity are
included in $Y_c$, which is complex. The Josephson inductance $L_J$ is
deduced from the measured junction resistance at room-temperature $R_T$, extrapolating it
down to the operating temperature~\cite{Gloos-2000} of $15\,{\rm mk}$ and using the
Ambegaokar-Baratoff relation, $E_J=h\Delta /(8e^2R_T)$. $C_J$ -- the
only free parameter -- is obtained by fitting the lowest three energy
levels of the numerical solution of the $\phi^6$ model to the measured spectrum~\cite{Paik-2011a}. Although $Y_c$ may in principle be obtained from
current-voltage measurements, this is not practical in this system,
where the antenna is hard to access non-invasively, being inside a closed high-Q cavity. Instead we use a finite element High Frequency
Simulation Software (HFSS) and obtain $Y_c(\omega)$ by
solving the Maxwell equations numerically. Details
on this simulation step are provided in the supplementary
material~\cite{supp-mat-BBQ}. 

From the zeros of the imaginary part of the admittance and their slopes we build and diagonalize the $\phi^6$
Hamiltonian in a truncated Hilbert space, keeping in total three
dressed modes
(one qubit and two cavity modes) and allowing for maximally ten
excitations per mode. The results of fitting the low-energy spectrum of six different samples are presented in Table~\ref{tab:1}, where we
also compare the predicted and measured qubit anharmonicities and
$\chi$-shifts. We find agreement with the measured spectrum at the sub-per cent
level and to within ten per cent with the measured anharmonicities and $\chi$-shifts.


{\em Conclusion and outlook.} We have presented a simple method to determine
the effective low-energy Hamiltonian of a wide class of superconducting circuits
containing lumped or distributed elements. This method is suitable for
weakly nonlinear circuits, for which the normal modes of the
linearized classical circuit provide a good basis in the quantum case. For an $N$ qubit system it requires only
the knowledge of an $N\times N$ (classical) impedance
matrix. By working in a basis
of dressed states, the parameters that appear in the Hamiltonian
incorporate much of the renormalization induced by the coupling
between a multi-level artificial atom and a multi-mode
resonator. Consequently, the number of free parameters is considerably
reduced as compared
with standard models based on the Jaynes-Cummings paradigm expressed in terms of the experimentally inaccessible bare parameters. We have demonstrated the
usefulness of this method in designing superconducting quantum
information processing units by computing the low-energy spectrum of a
3D-transmon. Finally, this model may represent a suitable starting point
for future investigations of the emerging ultra-strong coupling regime of cQED.

{\it Acknowledgments.} We thank Claudia De Grandi, Eustace Edwards and
Mazyar Mirrahimi for discussions and Mikhael Guy
from the Yale HPC center for support with numerical simulations. SEN
acknowledges financial support from the Swiss NSF. HP, GK, BV, LF, MD,
RS and SG acknowledge financial support from IARPA, ARO
(Contract W911NF-09-1-0514) and the
American NSF (Contract DMR-1004406). All statements of fact, opinion
or conclusions, contained herein are those of the authors and should
not be construed as representing the official views or policies of
IARPA, or the U.S. Government.
\bibliography{/home/sen/phys/library/bibtex/mybib.bib}

\begin{thebibliography}{33}%
\makeatletter
\providecommand \@ifxundefined [1]{%
 \@ifx{#1\undefined}
}%
\providecommand \@ifnum [1]{%
 \ifnum #1\expandafter \@firstoftwo
 \else \expandafter \@secondoftwo
 \fi
}%
\providecommand \@ifx [1]{%
 \ifx #1\expandafter \@firstoftwo
 \else \expandafter \@secondoftwo
 \fi
}%
\providecommand \natexlab [1]{#1}%
\providecommand \enquote  [1]{``#1''}%
\providecommand \bibnamefont  [1]{#1}%
\providecommand \bibfnamefont [1]{#1}%
\providecommand \citenamefont [1]{#1}%
\providecommand \href@noop [0]{\@secondoftwo}%
\providecommand \href [0]{\begingroup \@sanitize@url \@href}%
\providecommand \@href[1]{\@@startlink{#1}\@@href}%
\providecommand \@@href[1]{\endgroup#1\@@endlink}%
\providecommand \@sanitize@url [0]{\catcode `\\12\catcode `\$12\catcode
  `\&12\catcode `\#12\catcode `\^12\catcode `\_12\catcode `\%12\relax}%
\providecommand \@@startlink[1]{}%
\providecommand \@@endlink[0]{}%
\providecommand \url  [0]{\begingroup\@sanitize@url \@url }%
\providecommand \@url [1]{\endgroup\@href {#1}{\urlprefix }}%
\providecommand \urlprefix  [0]{URL }%
\providecommand \Eprint [0]{\href }%
\providecommand \doibase [0]{http://dx.doi.org/}%
\providecommand \selectlanguage [0]{\@gobble}%
\providecommand \bibinfo  [0]{\@secondoftwo}%
\providecommand \bibfield  [0]{\@secondoftwo}%
\providecommand \translation [1]{[#1]}%
\providecommand \BibitemOpen [0]{}%
\providecommand \bibitemStop [0]{}%
\providecommand \bibitemNoStop [0]{.\EOS\space}%
\providecommand \EOS [0]{\spacefactor3000\relax}%
\providecommand \BibitemShut  [1]{\csname bibitem#1\endcsname}%
\let\auto@bib@innerbib\@empty
\bibitem [{\citenamefont {Devoret}\ and\ \citenamefont
  {Martinis}(2004)}]{Devoret-2004a}%
  \BibitemOpen
  \bibfield  {author} {\bibinfo {author} {\bibfnamefont {M.~H.}\ \bibnamefont
  {Devoret}}\ and\ \bibinfo {author} {\bibfnamefont {J.~M.}\ \bibnamefont
  {Martinis}},\ }\href {\doibase 10.1007/s11128-004-3101-5} {\bibfield
  {journal} {\bibinfo  {journal} {Quantum Information Processing}\ }\textbf
  {\bibinfo {volume} {3}},\ \bibinfo {pages} {1} (\bibinfo {year}
  {2004})}\BibitemShut {NoStop}%
\bibitem [{\citenamefont {Wallraff}\ \emph {et~al.}(2004)\citenamefont
  {Wallraff}, , \citenamefont {Schuster}, \citenamefont {Blais}, \citenamefont
  {Frunzio}, \citenamefont {Huang}, \citenamefont {Majer}, \citenamefont
  {Kumar}, \citenamefont {Girvin},\ and\ \citenamefont
  {Schoelkopf}}]{Wallraff-2004a}%
  \BibitemOpen
  \bibfield  {author} {\bibinfo {author} {\bibfnamefont {A.}~\bibnamefont
  {Wallraff}}, , \bibinfo {author} {\bibfnamefont {D.~I.}\ \bibnamefont
  {Schuster}}, \bibinfo {author} {\bibfnamefont {A.}~\bibnamefont {Blais}},
  \bibinfo {author} {\bibfnamefont {L.}~\bibnamefont {Frunzio}}, \bibinfo
  {author} {\bibfnamefont {R.-S.}\ \bibnamefont {Huang}}, \bibinfo {author}
  {\bibfnamefont {J.}~\bibnamefont {Majer}}, \bibinfo {author} {\bibfnamefont
  {S.}~\bibnamefont {Kumar}}, \bibinfo {author} {\bibfnamefont {S.~M.}\
  \bibnamefont {Girvin}}, \ and\ \bibinfo {author} {\bibfnamefont {R.~J.}\
  \bibnamefont {Schoelkopf}},\ }\href {\doibase doi:10.1038/nature02851}
  {\bibfield  {journal} {\bibinfo  {journal} {Nature}\ }\textbf {\bibinfo
  {volume} {431}},\ \bibinfo {pages} {162} (\bibinfo {year}
  {2004})}\BibitemShut {NoStop}%
\bibitem [{\citenamefont {Paik}\ \emph {et~al.}(2011)\citenamefont {Paik},
  \citenamefont {Schuster}, \citenamefont {Bishop}, \citenamefont {Kirchmair},
  \citenamefont {Catelani}, \citenamefont {Sears}, \citenamefont {Johnson},
  \citenamefont {Reagor}, \citenamefont {Frunzio}, \citenamefont {Glazman},
  \citenamefont {Girvin}, \citenamefont {Devoret},\ and\ \citenamefont
  {Schoelkopf}}]{Paik-2011a}%
  \BibitemOpen
  \bibfield  {author} {\bibinfo {author} {\bibfnamefont {H.}~\bibnamefont
  {Paik}}, \bibinfo {author} {\bibfnamefont {D.~I.}\ \bibnamefont {Schuster}},
  \bibinfo {author} {\bibfnamefont {L.~S.}\ \bibnamefont {Bishop}}, \bibinfo
  {author} {\bibfnamefont {G.}~\bibnamefont {Kirchmair}}, \bibinfo {author}
  {\bibfnamefont {G.}~\bibnamefont {Catelani}}, \bibinfo {author}
  {\bibfnamefont {A.~P.}\ \bibnamefont {Sears}}, \bibinfo {author}
  {\bibfnamefont {B.~R.}\ \bibnamefont {Johnson}}, \bibinfo {author}
  {\bibfnamefont {M.~J.}\ \bibnamefont {Reagor}}, \bibinfo {author}
  {\bibfnamefont {L.}~\bibnamefont {Frunzio}}, \bibinfo {author} {\bibfnamefont
  {L.~I.}\ \bibnamefont {Glazman}}, \bibinfo {author} {\bibfnamefont {S.~M.}\
  \bibnamefont {Girvin}}, \bibinfo {author} {\bibfnamefont {M.~H.}\
  \bibnamefont {Devoret}}, \ and\ \bibinfo {author} {\bibfnamefont {R.~J.}\
  \bibnamefont {Schoelkopf}},\ }\href {\doibase 10.1103/PhysRevLett.107.240501}
  {\bibfield  {journal} {\bibinfo  {journal} {Phys. Rev. Lett.}\ }\textbf
  {\bibinfo {volume} {107}},\ \bibinfo {pages} {240501} (\bibinfo {year}
  {2011})}\BibitemShut {NoStop}%
\bibitem [{\citenamefont {Devoret}\ \emph {et~al.}(2007)\citenamefont
  {Devoret}, \citenamefont {Girvin},\ and\ \citenamefont
  {Schoelkopf}}]{Devoret-2007a}%
  \BibitemOpen
  \bibfield  {author} {\bibinfo {author} {\bibfnamefont {M.}~\bibnamefont
  {Devoret}}, \bibinfo {author} {\bibfnamefont {S.}~\bibnamefont {Girvin}}, \
  and\ \bibinfo {author} {\bibfnamefont {R.}~\bibnamefont {Schoelkopf}},\
  }\href@noop {} {\bibfield  {journal} {\bibinfo  {journal} {Ann. Phys}\
  }\textbf {\bibinfo {volume} {16}},\ \bibinfo {pages} {767} (\bibinfo {year}
  {2007})}\BibitemShut {NoStop}%
\bibitem [{\citenamefont {Houck}\ \emph {et~al.}(2008)\citenamefont {Houck},
  \citenamefont {Schreier}, \citenamefont {Johnson}, \citenamefont {Chow},
  \citenamefont {Koch}, \citenamefont {Gambetta}, \citenamefont {Schuster},
  \citenamefont {Frunzio}, \citenamefont {Devoret}, \citenamefont {Girvin},\
  and\ \citenamefont {Schoelkopf}}]{Houck-2008a}%
  \BibitemOpen
  \bibfield  {author} {\bibinfo {author} {\bibfnamefont {A.~A.}\ \bibnamefont
  {Houck}}, \bibinfo {author} {\bibfnamefont {J.~A.}\ \bibnamefont {Schreier}},
  \bibinfo {author} {\bibfnamefont {B.~R.}\ \bibnamefont {Johnson}}, \bibinfo
  {author} {\bibfnamefont {J.~M.}\ \bibnamefont {Chow}}, \bibinfo {author}
  {\bibfnamefont {J.}~\bibnamefont {Koch}}, \bibinfo {author} {\bibfnamefont
  {J.~M.}\ \bibnamefont {Gambetta}}, \bibinfo {author} {\bibfnamefont {D.~I.}\
  \bibnamefont {Schuster}}, \bibinfo {author} {\bibfnamefont {L.}~\bibnamefont
  {Frunzio}}, \bibinfo {author} {\bibfnamefont {M.~H.}\ \bibnamefont
  {Devoret}}, \bibinfo {author} {\bibfnamefont {S.~M.}\ \bibnamefont {Girvin}},
  \ and\ \bibinfo {author} {\bibfnamefont {R.~J.}\ \bibnamefont {Schoelkopf}},\
  }\href {\doibase 10.1103/PhysRevLett.101.080502} {\bibfield  {journal}
  {\bibinfo  {journal} {Phys. Rev. Lett.}\ }\textbf {\bibinfo {volume} {101}},\
  \bibinfo {pages} {080502} (\bibinfo {year} {2008})}\BibitemShut {NoStop}%
\bibitem [{\citenamefont {Bourassa}\ \emph {et~al.}(2009)\citenamefont
  {Bourassa}, \citenamefont {Gambetta}, \citenamefont {Abdumalikov},
  \citenamefont {Astafiev}, \citenamefont {Nakamura},\ and\ \citenamefont
  {Blais}}]{Bourassa-2009a}%
  \BibitemOpen
  \bibfield  {author} {\bibinfo {author} {\bibfnamefont {J.}~\bibnamefont
  {Bourassa}}, \bibinfo {author} {\bibfnamefont {J.~M.}\ \bibnamefont
  {Gambetta}}, \bibinfo {author} {\bibfnamefont {A.~A.}\ \bibnamefont
  {Abdumalikov}}, \bibinfo {author} {\bibfnamefont {O.}~\bibnamefont
  {Astafiev}}, \bibinfo {author} {\bibfnamefont {Y.}~\bibnamefont {Nakamura}},
  \ and\ \bibinfo {author} {\bibfnamefont {A.}~\bibnamefont {Blais}},\ }\href
  {\doibase 10.1103/PhysRevA.80.032109} {\bibfield  {journal} {\bibinfo
  {journal} {Phys. Rev. A}\ }\textbf {\bibinfo {volume} {80}},\ \bibinfo
  {pages} {032109} (\bibinfo {year} {2009})}\BibitemShut {NoStop}%
\bibitem [{\citenamefont {Niemczyk}\ \emph {et~al.}(2010)\citenamefont
  {Niemczyk}, \citenamefont {Deppe}, \citenamefont {Huebl}, \citenamefont
  {Menzel}, \citenamefont {Hocke}, \citenamefont {Schwarz}, \citenamefont
  {Garcia-Ripoll}, \citenamefont {D.~Zueco}, \citenamefont {Solano},
  \citenamefont {Marx},\ and\ \citenamefont {Gross}}]{Niemczyk-2010a}%
  \BibitemOpen
  \bibfield  {author} {\bibinfo {author} {\bibfnamefont {T.}~\bibnamefont
  {Niemczyk}}, \bibinfo {author} {\bibfnamefont {F.}~\bibnamefont {Deppe}},
  \bibinfo {author} {\bibfnamefont {H.}~\bibnamefont {Huebl}}, \bibinfo
  {author} {\bibfnamefont {E.~P.}\ \bibnamefont {Menzel}}, \bibinfo {author}
  {\bibfnamefont {F.}~\bibnamefont {Hocke}}, \bibinfo {author} {\bibfnamefont
  {M.~J.}\ \bibnamefont {Schwarz}}, \bibinfo {author} {\bibfnamefont {J.~J.}\
  \bibnamefont {Garcia-Ripoll}}, \bibinfo {author} {\bibfnamefont {T.~H.}\
  \bibnamefont {D.~Zueco}}, \bibinfo {author} {\bibfnamefont {E.}~\bibnamefont
  {Solano}}, \bibinfo {author} {\bibfnamefont {A.}~\bibnamefont {Marx}}, \ and\
  \bibinfo {author} {\bibfnamefont {R.}~\bibnamefont {Gross}},\ }\href
  {\doibase doi:10.1038/nphys1730} {\bibfield  {journal} {\bibinfo  {journal}
  {Nature Physics}\ }\textbf {\bibinfo {volume} {6}},\ \bibinfo {pages} {772}
  (\bibinfo {year} {2010})}\BibitemShut {NoStop}%
\bibitem [{\citenamefont {Filipp}\ \emph {et~al.}(2011)\citenamefont {Filipp},
  \citenamefont {G\"oppl}, \citenamefont {Fink}, \citenamefont {Baur},
  \citenamefont {Bianchetti}, \citenamefont {Steffen},\ and\ \citenamefont
  {Wallraff}}]{Filipp-2011a}%
  \BibitemOpen
  \bibfield  {author} {\bibinfo {author} {\bibfnamefont {S.}~\bibnamefont
  {Filipp}}, \bibinfo {author} {\bibfnamefont {M.}~\bibnamefont {G\"oppl}},
  \bibinfo {author} {\bibfnamefont {J.~M.}\ \bibnamefont {Fink}}, \bibinfo
  {author} {\bibfnamefont {M.}~\bibnamefont {Baur}}, \bibinfo {author}
  {\bibfnamefont {R.}~\bibnamefont {Bianchetti}}, \bibinfo {author}
  {\bibfnamefont {L.}~\bibnamefont {Steffen}}, \ and\ \bibinfo {author}
  {\bibfnamefont {A.}~\bibnamefont {Wallraff}},\ }\href {\doibase
  10.1103/PhysRevA.83.063827} {\bibfield  {journal} {\bibinfo  {journal} {Phys.
  Rev. A}\ }\textbf {\bibinfo {volume} {83}},\ \bibinfo {pages} {063827}
  (\bibinfo {year} {2011})}\BibitemShut {NoStop}%
\bibitem [{\citenamefont {Viehmann}\ \emph {et~al.}(2011)\citenamefont
  {Viehmann}, \citenamefont {von Delft},\ and\ \citenamefont
  {Marquardt}}]{Viehmann-2011a}%
  \BibitemOpen
  \bibfield  {author} {\bibinfo {author} {\bibfnamefont {O.}~\bibnamefont
  {Viehmann}}, \bibinfo {author} {\bibfnamefont {J.}~\bibnamefont {von Delft}},
  \ and\ \bibinfo {author} {\bibfnamefont {F.}~\bibnamefont {Marquardt}},\
  }\href {\doibase 10.1103/PhysRevLett.107.113602} {\bibfield  {journal}
  {\bibinfo  {journal} {Phys. Rev. Lett.}\ }\textbf {\bibinfo {volume} {107}},\
  \bibinfo {pages} {113602} (\bibinfo {year} {2011})}\BibitemShut {NoStop}%
\bibitem [{\citenamefont {Cottet}(2002)}]{ACottetThesis}%
  \BibitemOpen
  \bibfield  {author} {\bibinfo {author} {\bibfnamefont {A.}~\bibnamefont
  {Cottet}},\ }\emph {\bibinfo {title} {Implementation of a quantum bit in a
  superconducting circuit}},\ \href@noop {} {Ph.D. thesis},\ \bibinfo  {school}
  {Universit{\'e} Paris VI} (\bibinfo {year} {2002})\BibitemShut {NoStop}%
\bibitem [{\citenamefont {Koch}\ \emph {et~al.}(2007)\citenamefont {Koch},
  \citenamefont {Yu}, \citenamefont {Gambetta}, \citenamefont {Houck},
  \citenamefont {Schuster}, \citenamefont {Majer}, \citenamefont {Blais},
  \citenamefont {Devoret}, \citenamefont {Girvin},\ and\ \citenamefont
  {Schoelkopf}}]{Koch-2007a}%
  \BibitemOpen
  \bibfield  {author} {\bibinfo {author} {\bibfnamefont {J.}~\bibnamefont
  {Koch}}, \bibinfo {author} {\bibfnamefont {T.~M.}\ \bibnamefont {Yu}},
  \bibinfo {author} {\bibfnamefont {J.}~\bibnamefont {Gambetta}}, \bibinfo
  {author} {\bibfnamefont {A.~A.}\ \bibnamefont {Houck}}, \bibinfo {author}
  {\bibfnamefont {D.~I.}\ \bibnamefont {Schuster}}, \bibinfo {author}
  {\bibfnamefont {J.}~\bibnamefont {Majer}}, \bibinfo {author} {\bibfnamefont
  {A.}~\bibnamefont {Blais}}, \bibinfo {author} {\bibfnamefont {M.~H.}\
  \bibnamefont {Devoret}}, \bibinfo {author} {\bibfnamefont {S.~M.}\
  \bibnamefont {Girvin}}, \ and\ \bibinfo {author} {\bibfnamefont {R.~J.}\
  \bibnamefont {Schoelkopf}},\ }\href {\doibase 10.1103/PhysRevA.76.042319}
  {\bibfield  {journal} {\bibinfo  {journal} {Phys. Rev. A}\ }\textbf {\bibinfo
  {volume} {76}},\ \bibinfo {pages} {042319} (\bibinfo {year}
  {2007})}\BibitemShut {NoStop}%
\bibitem [{\citenamefont {Schreier}\ \emph {et~al.}(2008)\citenamefont
  {Schreier}, \citenamefont {Houck}, \citenamefont {Koch}, \citenamefont
  {Schuster}, \citenamefont {Johnson}, \citenamefont {Chow}, \citenamefont
  {Gambetta}, \citenamefont {Majer}, \citenamefont {Frunzio}, \citenamefont
  {Devoret}, \citenamefont {Girvin},\ and\ \citenamefont
  {Schoelkopf}}]{Schreier-2008a}%
  \BibitemOpen
  \bibfield  {author} {\bibinfo {author} {\bibfnamefont {J.~A.}\ \bibnamefont
  {Schreier}}, \bibinfo {author} {\bibfnamefont {A.~A.}\ \bibnamefont {Houck}},
  \bibinfo {author} {\bibfnamefont {J.}~\bibnamefont {Koch}}, \bibinfo {author}
  {\bibfnamefont {D.~I.}\ \bibnamefont {Schuster}}, \bibinfo {author}
  {\bibfnamefont {B.~R.}\ \bibnamefont {Johnson}}, \bibinfo {author}
  {\bibfnamefont {J.~M.}\ \bibnamefont {Chow}}, \bibinfo {author}
  {\bibfnamefont {J.~M.}\ \bibnamefont {Gambetta}}, \bibinfo {author}
  {\bibfnamefont {J.}~\bibnamefont {Majer}}, \bibinfo {author} {\bibfnamefont
  {L.}~\bibnamefont {Frunzio}}, \bibinfo {author} {\bibfnamefont {M.~H.}\
  \bibnamefont {Devoret}}, \bibinfo {author} {\bibfnamefont {S.~M.}\
  \bibnamefont {Girvin}}, \ and\ \bibinfo {author} {\bibfnamefont {R.~J.}\
  \bibnamefont {Schoelkopf}},\ }\href {\doibase 10.1103/PhysRevB.77.180502}
  {\bibfield  {journal} {\bibinfo  {journal} {Phys. Rev. B}\ }\textbf {\bibinfo
  {volume} {77}},\ \bibinfo {pages} {180502} (\bibinfo {year}
  {2008})}\BibitemShut {NoStop}%
\bibitem [{\citenamefont {Blais}\ \emph {et~al.}(2004)\citenamefont {Blais},
  \citenamefont {Huang}, \citenamefont {Wallraff}, \citenamefont {Girvin},\
  and\ \citenamefont {Schoelkopf}}]{Blais-2004a}%
  \BibitemOpen
  \bibfield  {author} {\bibinfo {author} {\bibfnamefont {A.}~\bibnamefont
  {Blais}}, \bibinfo {author} {\bibfnamefont {R.-S.}\ \bibnamefont {Huang}},
  \bibinfo {author} {\bibfnamefont {A.}~\bibnamefont {Wallraff}}, \bibinfo
  {author} {\bibfnamefont {S.~M.}\ \bibnamefont {Girvin}}, \ and\ \bibinfo
  {author} {\bibfnamefont {R.~J.}\ \bibnamefont {Schoelkopf}},\ }\href
  {\doibase 10.1103/PhysRevA.69.062320} {\bibfield  {journal} {\bibinfo
  {journal} {Phys. Rev. A}\ }\textbf {\bibinfo {volume} {69}},\ \bibinfo
  {pages} {062320} (\bibinfo {year} {2004})}\BibitemShut {NoStop}%
\bibitem [{\citenamefont {Jaynes}\ and\ \citenamefont
  {Cummings}(1963)}]{Jaynes-1963a}%
  \BibitemOpen
  \bibfield  {author} {\bibinfo {author} {\bibfnamefont {E.}~\bibnamefont
  {Jaynes}}\ and\ \bibinfo {author} {\bibfnamefont {F.}~\bibnamefont
  {Cummings}},\ }\href {\doibase 10.1109/PROC.1963.1664} {\bibfield  {journal}
  {\bibinfo  {journal} {Proceedings of the IEEE}\ }\textbf {\bibinfo {volume}
  {51}},\ \bibinfo {pages} {89 } (\bibinfo {year} {1963})}\BibitemShut
  {NoStop}%
\bibitem [{\citenamefont {Tavis}\ and\ \citenamefont
  {Cummings}(1968)}]{Tavis-1968a}%
  \BibitemOpen
  \bibfield  {author} {\bibinfo {author} {\bibfnamefont {M.}~\bibnamefont
  {Tavis}}\ and\ \bibinfo {author} {\bibfnamefont {F.~W.}\ \bibnamefont
  {Cummings}},\ }\href {\doibase 10.1103/PhysRev.170.379} {\bibfield  {journal}
  {\bibinfo  {journal} {Phys. Rev.}\ }\textbf {\bibinfo {volume} {170}},\
  \bibinfo {pages} {379} (\bibinfo {year} {1968})}\BibitemShut {NoStop}%
\bibitem [{\citenamefont {DiCarlo}\ \emph {et~al.}(2009)\citenamefont
  {DiCarlo}, \citenamefont {Chow}, \citenamefont {Gambetta}, \citenamefont
  {Bishop}, \citenamefont {Johnson}, \citenamefont {Schuster}, \citenamefont
  {Majer}, \citenamefont {Blais}, \citenamefont {Frunzio}, \citenamefont
  {Girvin},\ and\ \citenamefont {Schoelkopf}}]{DiCarlo-2009a}%
  \BibitemOpen
  \bibfield  {author} {\bibinfo {author} {\bibfnamefont {L.}~\bibnamefont
  {DiCarlo}}, \bibinfo {author} {\bibfnamefont {J.~M.}\ \bibnamefont {Chow}},
  \bibinfo {author} {\bibfnamefont {J.~M.}\ \bibnamefont {Gambetta}}, \bibinfo
  {author} {\bibfnamefont {L.~S.}\ \bibnamefont {Bishop}}, \bibinfo {author}
  {\bibfnamefont {B.~R.}\ \bibnamefont {Johnson}}, \bibinfo {author}
  {\bibfnamefont {D.~I.}\ \bibnamefont {Schuster}}, \bibinfo {author}
  {\bibfnamefont {J.}~\bibnamefont {Majer}}, \bibinfo {author} {\bibfnamefont
  {A.}~\bibnamefont {Blais}}, \bibinfo {author} {\bibfnamefont
  {L.}~\bibnamefont {Frunzio}}, \bibinfo {author} {\bibfnamefont {S.~M.}\
  \bibnamefont {Girvin}}, \ and\ \bibinfo {author} {\bibfnamefont {R.~J.}\
  \bibnamefont {Schoelkopf}},\ }\href {\doibase doi:10.1038/nature08121}
  {\bibfield  {journal} {\bibinfo  {journal} {Nature}\ }\textbf {\bibinfo
  {volume} {460}} (\bibinfo {year} {2009}),\
  doi:10.1038/nature08121}\BibitemShut {NoStop}%
\bibitem [{\citenamefont {Bourassa}\ and\ \citenamefont
  {Blais}()}]{Bourassa-2011-unpub}%
  \BibitemOpen
  \bibfield  {author} {\bibinfo {author} {\bibfnamefont {J.}~\bibnamefont
  {Bourassa}}\ and\ \bibinfo {author} {\bibfnamefont {A.}~\bibnamefont
  {Blais}},\ }\href@noop {} {}\bibinfo {note} {Private
  communication}\BibitemShut {NoStop}%
\bibitem [{Note1()}]{Note1}%
  \BibitemOpen
  \bibinfo {note} {In current realizations of the 3D-transmon qubits, the
  length of the antenna is between $1$ and $10\%$ of the wavelength of the
  fundamental bare cavity mode.}\BibitemShut {Stop}%
\bibitem [{\citenamefont {Reed}\ \emph {et~al.}(2012)\citenamefont {Reed},
  \citenamefont {DiCarlo}, \citenamefont {Nigg}, \citenamefont {Sun},
  \citenamefont {Frunzio}, \citenamefont {Girvin},\ and\ \citenamefont
  {Schoelkopf}}]{Reed-2012a}%
  \BibitemOpen
  \bibfield  {author} {\bibinfo {author} {\bibfnamefont {M.~D.}\ \bibnamefont
  {Reed}}, \bibinfo {author} {\bibfnamefont {L.}~\bibnamefont {DiCarlo}},
  \bibinfo {author} {\bibfnamefont {S.~E.}\ \bibnamefont {Nigg}}, \bibinfo
  {author} {\bibfnamefont {L.}~\bibnamefont {Sun}}, \bibinfo {author}
  {\bibfnamefont {L.}~\bibnamefont {Frunzio}}, \bibinfo {author} {\bibfnamefont
  {S.~M.}\ \bibnamefont {Girvin}}, \ and\ \bibinfo {author} {\bibfnamefont
  {R.~J.}\ \bibnamefont {Schoelkopf}},\ }\href {\doibase
  doi:10.1038/nature10786} {\bibfield  {journal} {\bibinfo  {journal} {Nature}\
  }\textbf {\bibinfo {volume} {482}},\ \bibinfo {pages} {382} (\bibinfo {year}
  {2012})}\BibitemShut {NoStop}%
\bibitem [{\citenamefont {Manucharyan}\ \emph {et~al.}(2007)\citenamefont
  {Manucharyan}, \citenamefont {Boaknin}, \citenamefont {Metcalfe},
  \citenamefont {Vijay}, \citenamefont {Siddiqi},\ and\ \citenamefont
  {Devoret}}]{Manucharyan-2007a}%
  \BibitemOpen
  \bibfield  {author} {\bibinfo {author} {\bibfnamefont {V.~E.}\ \bibnamefont
  {Manucharyan}}, \bibinfo {author} {\bibfnamefont {E.}~\bibnamefont
  {Boaknin}}, \bibinfo {author} {\bibfnamefont {M.}~\bibnamefont {Metcalfe}},
  \bibinfo {author} {\bibfnamefont {R.}~\bibnamefont {Vijay}}, \bibinfo
  {author} {\bibfnamefont {I.}~\bibnamefont {Siddiqi}}, \ and\ \bibinfo
  {author} {\bibfnamefont {M.}~\bibnamefont {Devoret}},\ }\href {\doibase
  10.1103/PhysRevB.76.014524} {\bibfield  {journal} {\bibinfo  {journal} {Phys.
  Rev. B}\ }\textbf {\bibinfo {volume} {76}},\ \bibinfo {pages} {014524}
  (\bibinfo {year} {2007})}\BibitemShut {NoStop}%
\bibitem [{\citenamefont {Foster}(1924)}]{Foster-1924}%
  \BibitemOpen
  \bibfield  {author} {\bibinfo {author} {\bibfnamefont {R.~M.}\ \bibnamefont
  {Foster}},\ }\href@noop {} {\bibfield  {journal} {\bibinfo  {journal} {Bell
  System Technical Journal}\ }\textbf {\bibinfo {volume} {3}},\ \bibinfo
  {pages} {260} (\bibinfo {year} {1924})}\BibitemShut {NoStop}%
\bibitem [{\citenamefont {Beinger}\ \emph {et~al.}(1945)\citenamefont
  {Beinger}, \citenamefont {Dicke}, \citenamefont {Marcuvitz}, \citenamefont
  {Montgomery},\ and\ \citenamefont {Purcell}}]{PMC-1948}%
  \BibitemOpen
  \bibfield  {author} {\bibinfo {author} {\bibfnamefont {E.~R.}\ \bibnamefont
  {Beinger}}, \bibinfo {author} {\bibfnamefont {R.~H.}\ \bibnamefont {Dicke}},
  \bibinfo {author} {\bibfnamefont {N.}~\bibnamefont {Marcuvitz}}, \bibinfo
  {author} {\bibfnamefont {C.~G.}\ \bibnamefont {Montgomery}}, \ and\ \bibinfo
  {author} {\bibfnamefont {E.~M.}\ \bibnamefont {Purcell}},\ }\href@noop {}
  {\emph {\bibinfo {title} {Principles of Microwave Circuits}}},\ edited by\
  \bibinfo {editor} {\bibfnamefont {C.~G.}\ \bibnamefont {Montgomery}},
  \bibinfo {editor} {\bibfnamefont {R.~H.}\ \bibnamefont {Dicke}}, \ and\
  \bibinfo {editor} {\bibfnamefont {E.~M.}\ \bibnamefont {Purcell}}\ (\bibinfo
  {publisher} {MIT Radiation Laboratory},\ \bibinfo {year} {1945})\BibitemShut
  {NoStop}%
\bibitem [{Note2()}]{Note2}%
  \BibitemOpen
  \bibinfo {note} {The case of infinitely many discrete modes necessitates an
  extension of Foster's theorem as discussed in~\cite {Zinn-1951a}, but the
  results presented here still apply.}\BibitemShut {Stop}%
\bibitem [{\citenamefont {Devoret}(1995)}]{Devoret-1995a}%
  \BibitemOpen
  \bibfield  {author} {\bibinfo {author} {\bibfnamefont {M.~H.}\ \bibnamefont
  {Devoret}},\ }\enquote {\bibinfo {title} {Quantum fluctuations in electrical
  circuits},}\ \ (\bibinfo  {publisher} {Elsevier Science B. V.},\ \bibinfo
  {year} {1995})\ Chap.~\bibinfo {chapter} {10}, p.\ \bibinfo {pages} {351},\
  \bibinfo {note} {les Houches, Session LXIII}\BibitemShut {NoStop}%
\bibitem [{\citenamefont {Clerk}\ \emph {et~al.}(2010)\citenamefont {Clerk},
  \citenamefont {Devoret}, \citenamefont {Girvin}, \citenamefont {Marquardt},\
  and\ \citenamefont {Schoelkopf}}]{Clerk-2010a}%
  \BibitemOpen
  \bibfield  {author} {\bibinfo {author} {\bibfnamefont {A.~A.}\ \bibnamefont
  {Clerk}}, \bibinfo {author} {\bibfnamefont {M.~H.}\ \bibnamefont {Devoret}},
  \bibinfo {author} {\bibfnamefont {S.~M.}\ \bibnamefont {Girvin}}, \bibinfo
  {author} {\bibfnamefont {F.}~\bibnamefont {Marquardt}}, \ and\ \bibinfo
  {author} {\bibfnamefont {R.~J.}\ \bibnamefont {Schoelkopf}},\ }\href
  {\doibase 10.1103/RevModPhys.82.1155} {\bibfield  {journal} {\bibinfo
  {journal} {Rev. Mod. Phys.}\ }\textbf {\bibinfo {volume} {82}},\ \bibinfo
  {pages} {1155} (\bibinfo {year} {2010})}\BibitemShut {NoStop}%
\bibitem [{\citenamefont {Boissonneault}\ \emph {et~al.}(2010)\citenamefont
  {Boissonneault}, \citenamefont {Gambetta},\ and\ \citenamefont
  {Blais}}]{Boissonneault-2010a}%
  \BibitemOpen
  \bibfield  {author} {\bibinfo {author} {\bibfnamefont {M.}~\bibnamefont
  {Boissonneault}}, \bibinfo {author} {\bibfnamefont {J.~M.}\ \bibnamefont
  {Gambetta}}, \ and\ \bibinfo {author} {\bibfnamefont {A.}~\bibnamefont
  {Blais}},\ }\href {\doibase 10.1103/PhysRevLett.105.100504} {\bibfield
  {journal} {\bibinfo  {journal} {Phys. Rev. Lett.}\ }\textbf {\bibinfo
  {volume} {105}},\ \bibinfo {pages} {100504} (\bibinfo {year}
  {2010})}\BibitemShut {NoStop}%
\bibitem [{\citenamefont {Caldeira}\ and\ \citenamefont
  {Leggett}(1981)}]{Caldeira-1981a}%
  \BibitemOpen
  \bibfield  {author} {\bibinfo {author} {\bibfnamefont {A.~O.}\ \bibnamefont
  {Caldeira}}\ and\ \bibinfo {author} {\bibfnamefont {A.~J.}\ \bibnamefont
  {Leggett}},\ }\href {\doibase 10.1103/PhysRevLett.46.211} {\bibfield
  {journal} {\bibinfo  {journal} {Phys. Rev. Lett.}\ }\textbf {\bibinfo
  {volume} {46}},\ \bibinfo {pages} {211} (\bibinfo {year} {1981})}\BibitemShut
  {NoStop}%
\bibitem [{sup()}]{supp-mat-BBQ}%
  \BibitemOpen
  \href@noop {} {}\bibinfo {note} {See appended supplementary
  material.}\BibitemShut {Stop}%
\bibitem [{\citenamefont {Gloos}\ \emph {et~al.}(2000)\citenamefont {Gloos},
  \citenamefont {Poikolainen},\ and\ \citenamefont {Pekola}}]{Gloos-2000}%
  \BibitemOpen
  \bibfield  {author} {\bibinfo {author} {\bibfnamefont {K.}~\bibnamefont
  {Gloos}}, \bibinfo {author} {\bibfnamefont {R.~S.}\ \bibnamefont
  {Poikolainen}}, \ and\ \bibinfo {author} {\bibfnamefont {J.~P.}\ \bibnamefont
  {Pekola}},\ }\href {\doibase 10.1063/1.1320861} {\bibfield  {journal}
  {\bibinfo  {journal} {Applied Physics Letters}\ }\textbf {\bibinfo {volume}
  {77}},\ \bibinfo {pages} {2915} (\bibinfo {year} {2000})}\BibitemShut
  {NoStop}%
\bibitem [{\citenamefont {Zinn}(1951)}]{Zinn-1951a}%
  \BibitemOpen
  \bibfield  {author} {\bibinfo {author} {\bibfnamefont {M.~K.}\ \bibnamefont
  {Zinn}},\ }\href@noop {} {\bibfield  {journal} {\bibinfo  {journal} {Bell
  System Technical Journal}\ }\textbf {\bibinfo {volume} {31}},\ \bibinfo
  {pages} {378} (\bibinfo {year} {1951})}\BibitemShut {NoStop}%
\bibitem [{\citenamefont {Pobel}(1937)}]{Pobel-1937}%
  \BibitemOpen
  \bibfield  {author} {\bibinfo {author} {\bibfnamefont {F.}~\bibnamefont
  {Pobel}},\ }\href@noop {} {\emph {\bibinfo {title} {Matter and Methods at Low
  Temperatures}}},\ \bibinfo {edition} {3rd}\ ed.\ (\bibinfo  {publisher}
  {Springer},\ \bibinfo {year} {1937})\BibitemShut {NoStop}%
\bibitem [{\citenamefont {Krupka}\ \emph {et~al.}(1999)\citenamefont {Krupka},
  \citenamefont {Derzakowski}, \citenamefont {Tobar}, \citenamefont
  {Hartnett},\ and\ \citenamefont {Geyer}}]{Krupka-1999}%
  \BibitemOpen
  \bibfield  {author} {\bibinfo {author} {\bibfnamefont {J.}~\bibnamefont
  {Krupka}}, \bibinfo {author} {\bibfnamefont {K.}~\bibnamefont {Derzakowski}},
  \bibinfo {author} {\bibfnamefont {M.}~\bibnamefont {Tobar}}, \bibinfo
  {author} {\bibfnamefont {J.}~\bibnamefont {Hartnett}}, \ and\ \bibinfo
  {author} {\bibfnamefont {R.~G.}\ \bibnamefont {Geyer}},\ }\href
  {http://stacks.iop.org/0957-0233/10/i=5/a=308} {\bibfield  {journal}
  {\bibinfo  {journal} {Measurement Science and Technology}\ }\textbf {\bibinfo
  {volume} {10}},\ \bibinfo {pages} {387} (\bibinfo {year} {1999})}\BibitemShut
  {NoStop}%
\bibitem [{Note3()}]{Note3}%
  \BibitemOpen
  \bibinfo {note} {Note that strictly speaking the commutator is rather
  $[\protect \qopname \relax o{exp}(i\varphi _s),n_s]=-\hbar \protect \qopname
  \relax o{exp}(i\varphi _s)$, but as we neglect charge dispersion, it is
  consistent to neglect the $2\pi $-periodicity of the commutation
  relation.}\BibitemShut {Stop}%
\end{thebibliography}%


%


\begin{thebibliography}{7}%
\makeatletter
\providecommand \@ifxundefined [1]{%
 \@ifx{#1\undefined}
}%
\providecommand \@ifnum [1]{%
 \ifnum #1\expandafter \@firstoftwo
 \else \expandafter \@secondoftwo
 \fi
}%
\providecommand \@ifx [1]{%
 \ifx #1\expandafter \@firstoftwo
 \else \expandafter \@secondoftwo
 \fi
}%
\providecommand \natexlab [1]{#1}%
\providecommand \enquote  [1]{``#1''}%
\providecommand \bibnamefont  [1]{#1}%
\providecommand \bibfnamefont [1]{#1}%
\providecommand \citenamefont [1]{#1}%
\providecommand \href@noop [0]{\@secondoftwo}%
\providecommand \href [0]{\begingroup \@sanitize@url \@href}%
\providecommand \@href[1]{\@@startlink{#1}\@@href}%
\providecommand \@@href[1]{\endgroup#1\@@endlink}%
\providecommand \@sanitize@url [0]{\catcode `\\12\catcode `\$12\catcode
  `\&12\catcode `\#12\catcode `\^12\catcode `\_12\catcode `\%12\relax}%
\providecommand \@@startlink[1]{}%
\providecommand \@@endlink[0]{}%
\providecommand \url  [0]{\begingroup\@sanitize@url \@url }%
\providecommand \@url [1]{\endgroup\@href {#1}{\urlprefix }}%
\providecommand \urlprefix  [0]{URL }%
\providecommand \Eprint [0]{\href }%
\providecommand \doibase [0]{http://dx.doi.org/}%
\providecommand \selectlanguage [0]{\@gobble}%
\providecommand \bibinfo  [0]{\@secondoftwo}%
\providecommand \bibfield  [0]{\@secondoftwo}%
\providecommand \translation [1]{[#1]}%
\providecommand \BibitemOpen [0]{}%
\providecommand \bibitemStop [0]{}%
\providecommand \bibitemNoStop [0]{.\EOS\space}%
\providecommand \EOS [0]{\spacefactor3000\relax}%
\providecommand \BibitemShut  [1]{\csname bibitem#1\endcsname}%
\let\auto@bib@innerbib\@empty
\bibitem [{\citenamefont {Pobel}(1937)}]{Pobel-1937}%
  \BibitemOpen
  \bibfield  {author} {\bibinfo {author} {\bibfnamefont {F.}~\bibnamefont
  {Pobel}},\ }\href@noop {} {\emph {\bibinfo {title} {Matter and Methods at Low
  Temperatures}}},\ \bibinfo {edition} {3rd}\ ed.\ (\bibinfo  {publisher}
  {Springer},\ \bibinfo {year} {1937})\BibitemShut {NoStop}%
\bibitem [{\citenamefont {Krupka}\ \emph {et~al.}(1999)\citenamefont {Krupka},
  \citenamefont {Derzakowski}, \citenamefont {Tobar}, \citenamefont
  {Hartnett},\ and\ \citenamefont {Geyer}}]{Krupka-1999}%
  \BibitemOpen
  \bibfield  {author} {\bibinfo {author} {\bibfnamefont {J.}~\bibnamefont
  {Krupka}}, \bibinfo {author} {\bibfnamefont {K.}~\bibnamefont {Derzakowski}},
  \bibinfo {author} {\bibfnamefont {M.}~\bibnamefont {Tobar}}, \bibinfo
  {author} {\bibfnamefont {J.}~\bibnamefont {Hartnett}}, \ and\ \bibinfo
  {author} {\bibfnamefont {R.~G.}\ \bibnamefont {Geyer}},\ }\href
  {http://stacks.iop.org/0957-0233/10/i=5/a=308} {\bibfield  {journal}
  {\bibinfo  {journal} {Measurement Science and Technology}\ }\textbf {\bibinfo
  {volume} {10}},\ \bibinfo {pages} {387} (\bibinfo {year} {1999})}\BibitemShut
  {NoStop}%
\bibitem [{\citenamefont {Paik}\ \emph {et~al.}(2011)\citenamefont {Paik},
  \citenamefont {Schuster}, \citenamefont {Bishop}, \citenamefont {Kirchmair},
  \citenamefont {Catelani}, \citenamefont {Sears}, \citenamefont {Johnson},
  \citenamefont {Reagor}, \citenamefont {Frunzio}, \citenamefont {Glazman},
  \citenamefont {Girvin}, \citenamefont {Devoret},\ and\ \citenamefont
  {Schoelkopf}}]{Paik-2011a}%
  \BibitemOpen
  \bibfield  {author} {\bibinfo {author} {\bibfnamefont {H.}~\bibnamefont
  {Paik}}, \bibinfo {author} {\bibfnamefont {D.~I.}\ \bibnamefont {Schuster}},
  \bibinfo {author} {\bibfnamefont {L.~S.}\ \bibnamefont {Bishop}}, \bibinfo
  {author} {\bibfnamefont {G.}~\bibnamefont {Kirchmair}}, \bibinfo {author}
  {\bibfnamefont {G.}~\bibnamefont {Catelani}}, \bibinfo {author}
  {\bibfnamefont {A.~P.}\ \bibnamefont {Sears}}, \bibinfo {author}
  {\bibfnamefont {B.~R.}\ \bibnamefont {Johnson}}, \bibinfo {author}
  {\bibfnamefont {M.~J.}\ \bibnamefont {Reagor}}, \bibinfo {author}
  {\bibfnamefont {L.}~\bibnamefont {Frunzio}}, \bibinfo {author} {\bibfnamefont
  {L.~I.}\ \bibnamefont {Glazman}}, \bibinfo {author} {\bibfnamefont {S.~M.}\
  \bibnamefont {Girvin}}, \bibinfo {author} {\bibfnamefont {M.~H.}\
  \bibnamefont {Devoret}}, \ and\ \bibinfo {author} {\bibfnamefont {R.~J.}\
  \bibnamefont {Schoelkopf}},\ }\href {\doibase 10.1103/PhysRevLett.107.240501}
  {\bibfield  {journal} {\bibinfo  {journal} {Phys. Rev. Lett.}\ }\textbf
  {\bibinfo {volume} {107}},\ \bibinfo {pages} {240501} (\bibinfo {year}
  {2011})}\BibitemShut {NoStop}%
\bibitem [{Note1()}]{Note1}%
  \BibitemOpen
  \bibinfo {note} {Note that strictly speaking the commutator is rather
  $[\protect \qopname \relax o{exp}(i\varphi _n),n_n]=-\hbar \protect \qopname
  \relax o{exp}(i\varphi _n)$, but as we neglect charge dispersion, it is
  consistent to neglect the $2\pi $-periodicity of the commutation
  relation.}\BibitemShut {Stop}%
\bibitem [{\citenamefont {Foster}(1924)}]{Foster-1924}%
  \BibitemOpen
  \bibfield  {author} {\bibinfo {author} {\bibfnamefont {R.~M.}\ \bibnamefont
  {Foster}},\ }\href@noop {} {\bibfield  {journal} {\bibinfo  {journal} {Bell
  System Technical Journal}\ }\textbf {\bibinfo {volume} {3}},\ \bibinfo
  {pages} {260} (\bibinfo {year} {1924})}\BibitemShut {NoStop}%
\bibitem [{\citenamefont {Devoret}(1995)}]{Devoret-1995a}%
  \BibitemOpen
  \bibfield  {author} {\bibinfo {author} {\bibfnamefont {M.~H.}\ \bibnamefont
  {Devoret}},\ }\enquote {\bibinfo {title} {Quantum fluctuations in electrical
  circuits},}\ \ (\bibinfo  {publisher} {Elsevier Science B. V.},\ \bibinfo
  {year} {1995})\ Chap.~\bibinfo {chapter} {10}, p.\ \bibinfo {pages} {351},\
  \bibinfo {note} {les Houches, Session LXIII}\BibitemShut {NoStop}%
\bibitem [{\citenamefont {Clerk}\ \emph {et~al.}(2010)\citenamefont {Clerk},
  \citenamefont {Devoret}, \citenamefont {Girvin}, \citenamefont {Marquardt},\
  and\ \citenamefont {Schoelkopf}}]{Clerk-2010a}%
  \BibitemOpen
  \bibfield  {author} {\bibinfo {author} {\bibfnamefont {A.~A.}\ \bibnamefont
  {Clerk}}, \bibinfo {author} {\bibfnamefont {M.~H.}\ \bibnamefont {Devoret}},
  \bibinfo {author} {\bibfnamefont {S.~M.}\ \bibnamefont {Girvin}}, \bibinfo
  {author} {\bibfnamefont {F.}~\bibnamefont {Marquardt}}, \ and\ \bibinfo
  {author} {\bibfnamefont {R.~J.}\ \bibnamefont {Schoelkopf}},\ }\href
  {\doibase 10.1103/RevModPhys.82.1155} {\bibfield  {journal} {\bibinfo
  {journal} {Rev. Mod. Phys.}\ }\textbf {\bibinfo {volume} {82}},\ \bibinfo
  {pages} {1155} (\bibinfo {year} {2010})}\BibitemShut {NoStop}%
\end{thebibliography}%

\cleardoublepage
\newpage

\begin{widetext}
\setlength{\textwidth}{15cm}
\setlength{\oddsidemargin}{0.7cm}
\setlength{\evensidemargin}{0.7cm}
\begin{center}{\Large\bf Supplementary Material for\\``Black-box superconducting circuit quantization'' }\end{center}
\vspace{0.2cm}
\begin{center}Simon E. Nigg, Hanhee Paik, Brian
  Vlastakis, Gerhard Kirchmair, Shyam~Shankar\\Luigi~Frunzio, Michel Devoret,
  Robert Schoelkopf and Steven Girvin\\
{\it Departments of Physics and Applied Physics, Yale University, New Haven, CT
  06520, USA}\\(Dated: \today)
\end{center}
\begin{center}\begin{minipage}[t]{0.8\textwidth}
\hspace{0.3cm}\small{These notes provide further details on the HFSS simulation of the
cavity admittance used to build the effective low-energy Hamiltonian
in the black-box quantization approach to compare with the single
junction experiment and on the black-box quantization
method for the multi-qubit case.}
\end{minipage}
\end{center}
\section{HFSS modeling of a 3D-transmon}

As discussed in the main text, the information about the spectrum of
the quantum circuit, is encoded in the admittance at the port of
the Josephson junction $Y(\omega)=Z(\omega)^{-1}$. More precisely, it
is sufficient to know the real roots and the
derivative of $Y$ at these points.

Assuming that the size of the junction
is negligibly small compared with the wavelength of the
lower modes of the electromagnetic field in the cavity, it is
appropriate to approximate the admittance of the linear part of the junction by a simple
lumped element parallel LC oscillator with inductance $L_J$ and
capacitance $C_J$ in parallel with the rest of the linear resonator. Hence the admittance can be decomposed as
\begin{equation}
Y(\omega) = j\omega C_J-\frac{j}{\omega L_J}+Y_c(\omega)\,,
\end{equation}
where $Y_c(\omega)$ is the admittance of the system without the
junction. The latter quantity can in principle be directly measured
but in this particular design a measurement is not practical. Instead
we simulate the classical system {\em without} the junction by solving
Maxwell's equations numerically using HFSS. Fig.~\ref{fig:3b} shows a graphical representation of the different meshes used to
represent the different elements of the cavity and antenna system. The
smaller the element, the finer the mesh needs to be for accuracy
and convergence.
In
\begin{figure}
\includegraphics[width=0.5\textwidth]{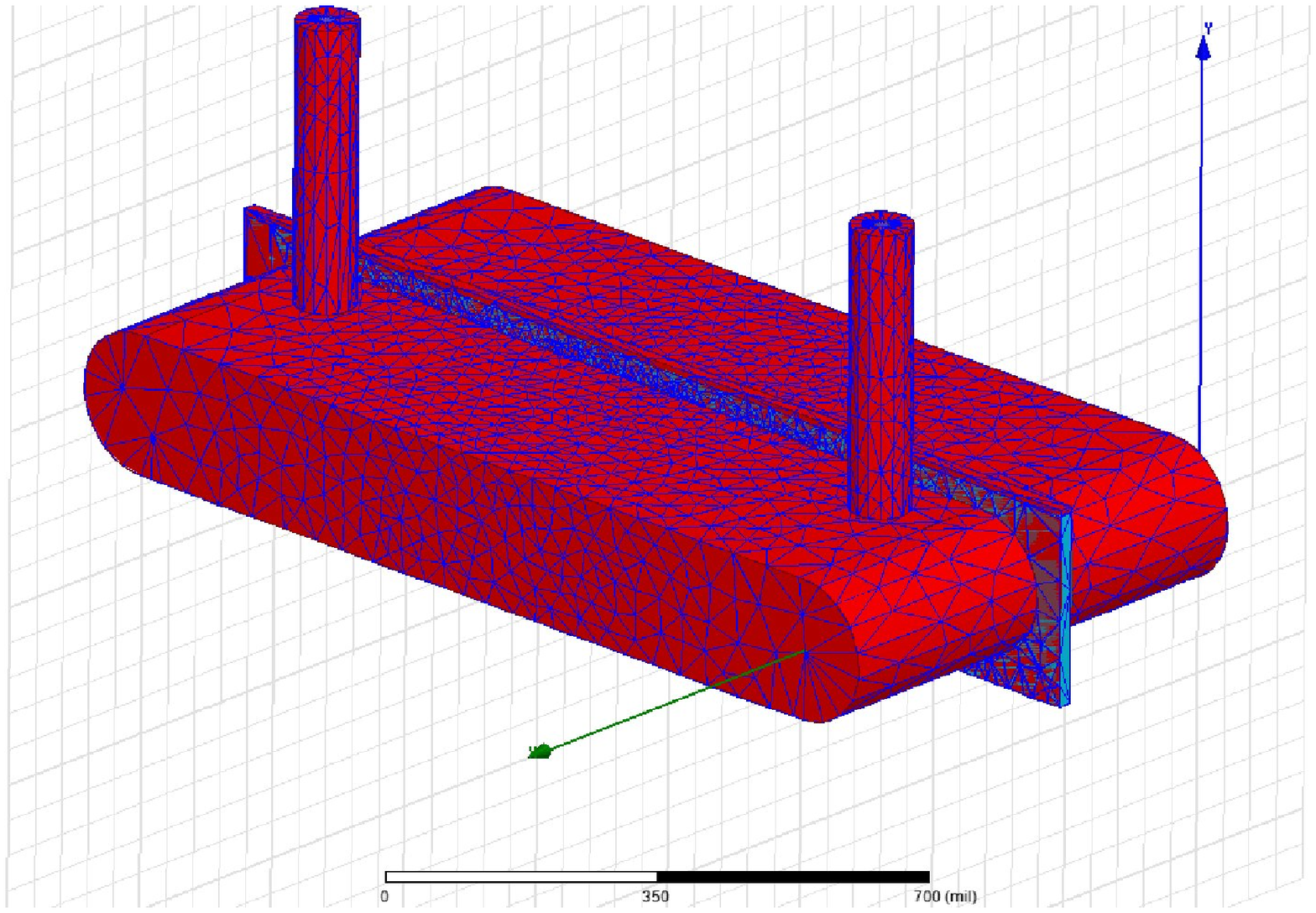}\hfill\includegraphics[width=0.5\textwidth]{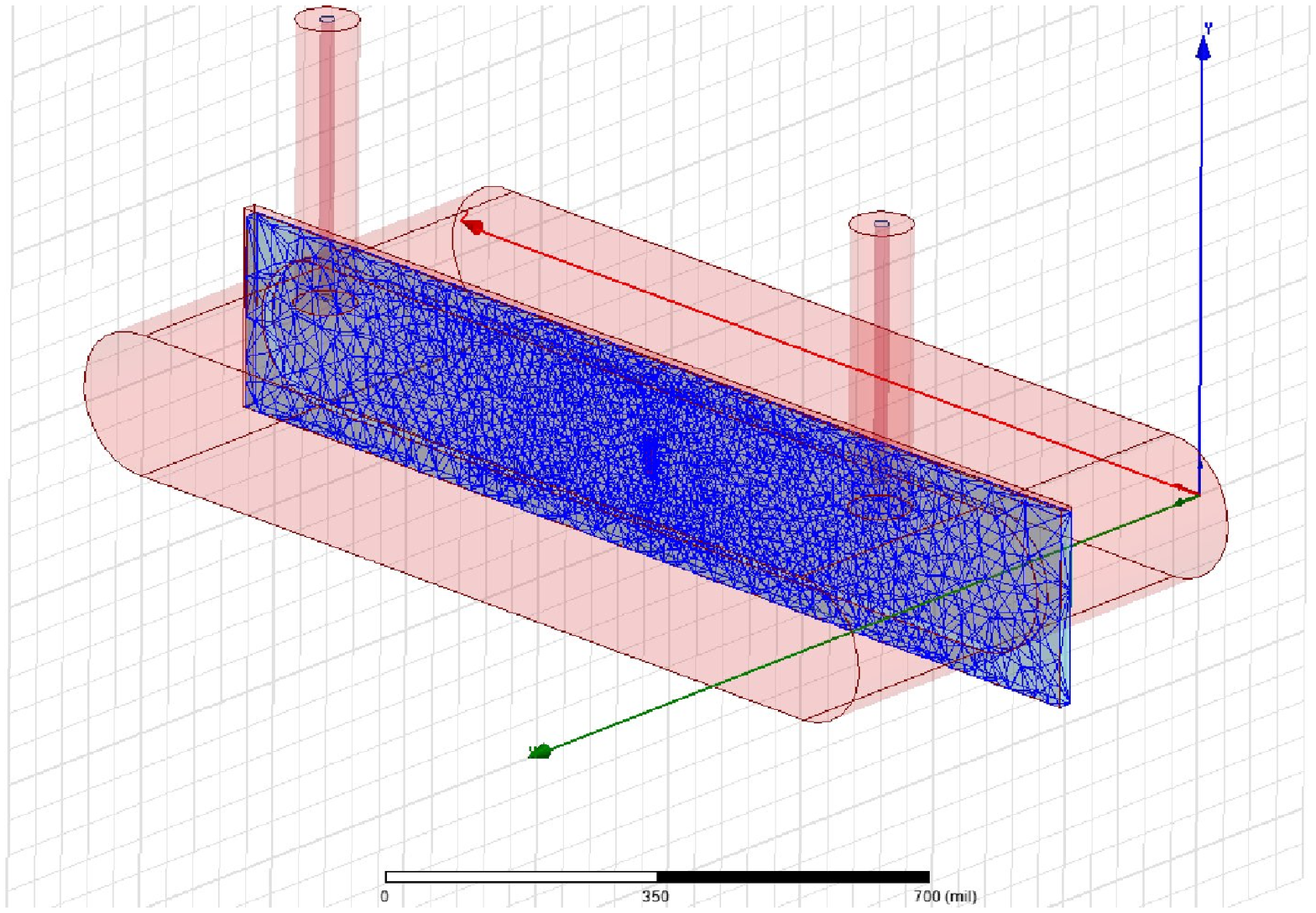}\\
\includegraphics[width=0.5\textwidth]{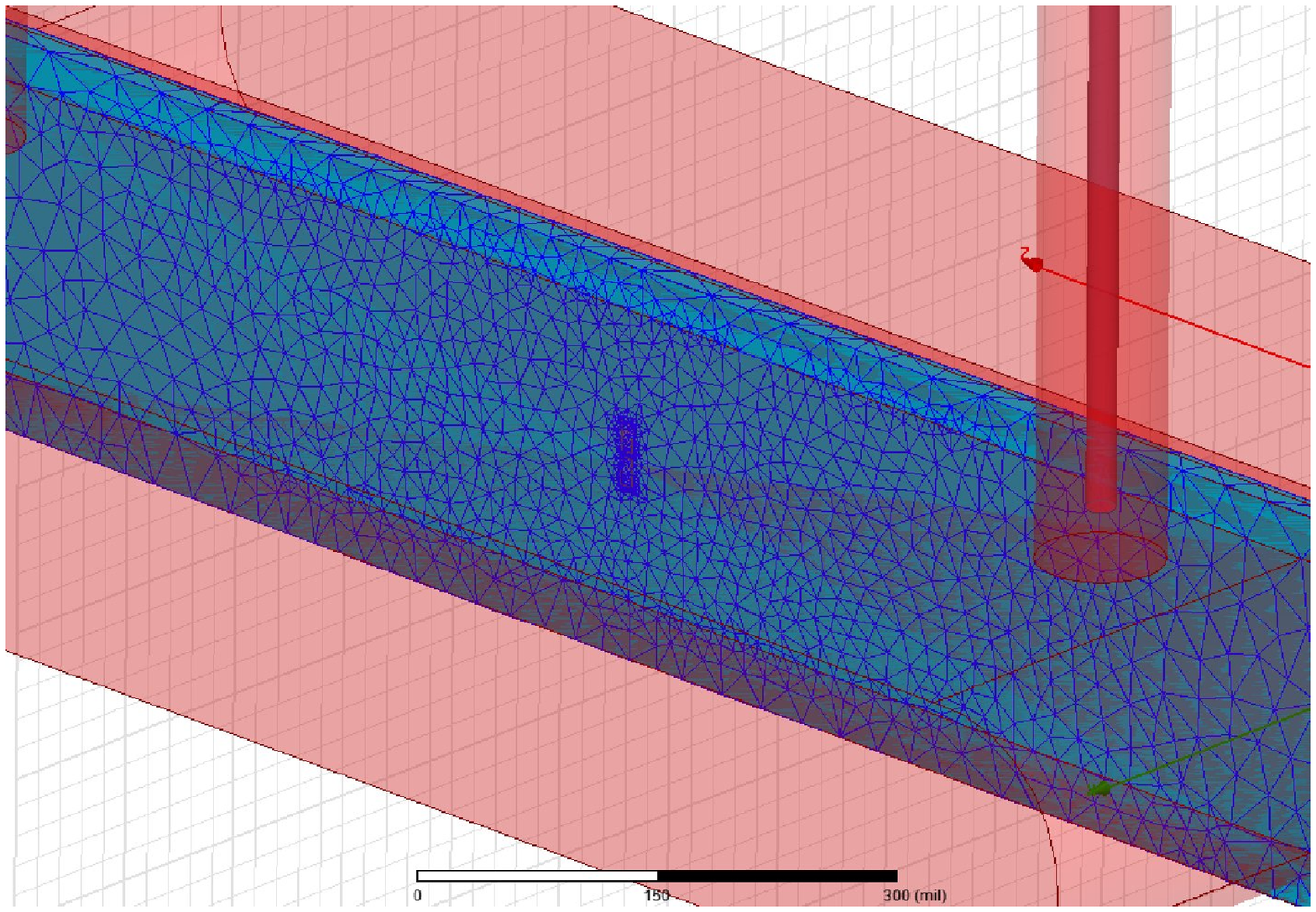}\hfill\includegraphics[width=0.5\textwidth]{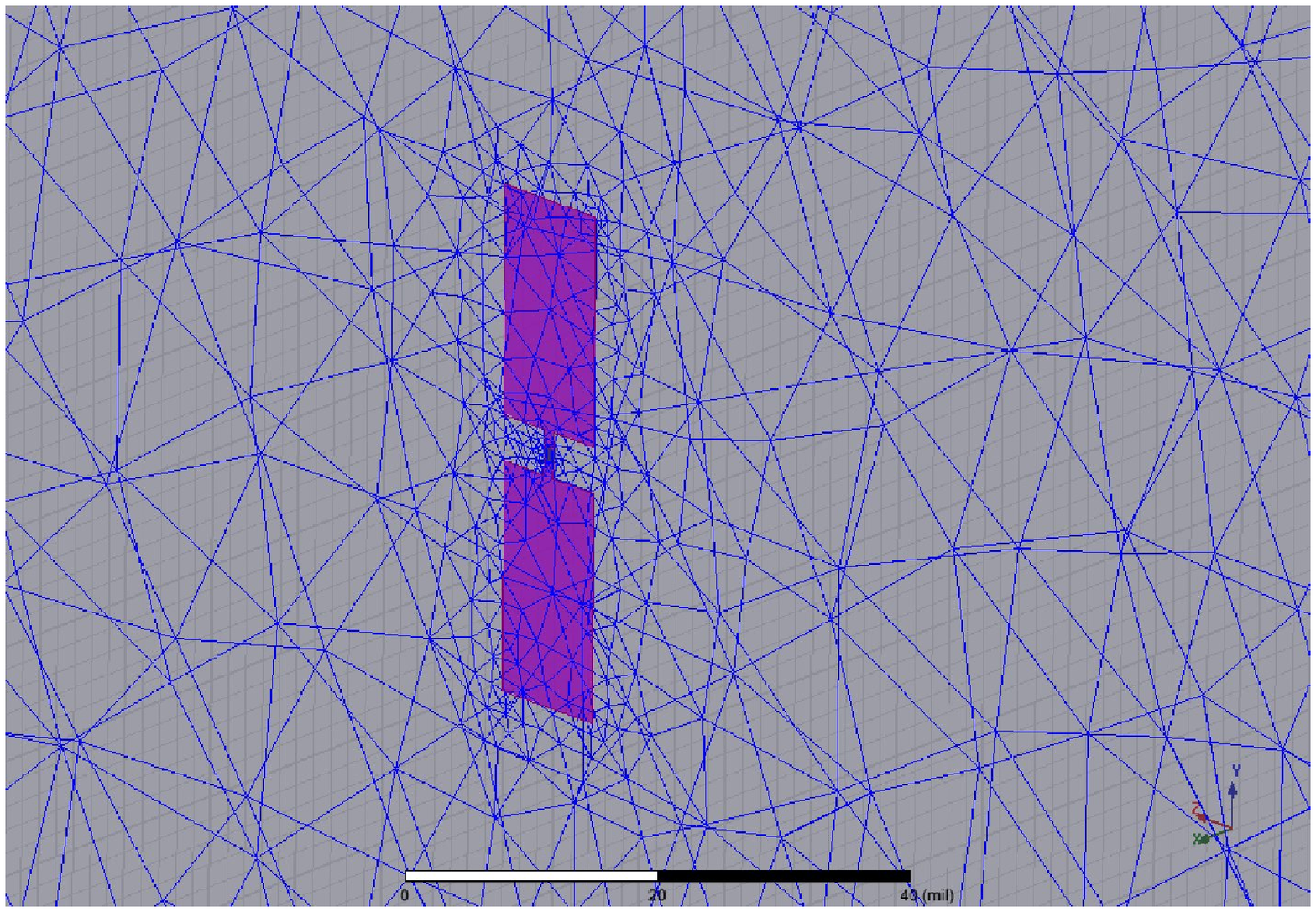}\caption{(Color
  online) HFSS
  model of a 3D-transmon. (a) The 3D resonator with input and output
  ports. These are terminated by $50\,{\rm Ohm}$ ports. (b)
  Transparent view of the cavity showing the sapphire
  substrate. Because the electric field is concentrated in the
  dielectric, a finer mesh is used. (c) and (d) Zoom-ins on the antenna placed
  on top of the substrate. The mesh is finest around the antenna.\label{fig:3b}}
\end{figure}
this finite element
simulation all metallic parts (Antenna and
cavity boundaries made of pure aluminum), are treated as
perfect conductors with zero resistance. In doing so, we neglect the
kinetic inductance of the antenna and cavity. The finite
London penetration depth of roughly $\lambda\approx 15\,{\rm nm}$
would lead to an effective increase of the cavity size and hence a
decrease of the cavity frequency of about $10\,{\rm kHz}$. Furthermore,
the kinetic inductance of the antenna and wire connecting the two
antenna pads to the Josephson junction can be estimated as
\begin{equation}
L_k=\frac{\lambda\mu_0}{2\tanh\left(\frac{d}{2\lambda}\right)}\left[\frac{L}{W}+\frac{l}{w}\right]\,,
\end{equation} 
where $d\approx 100\,{\rm nm}$ is the thickness of the aluminum layer,
$L\approx 1\,{\rm mm}$ is the total length and $W\approx 250\,{\rm
  \mu m}$ the width of the antenna and $l\approx 34\,{\rm \mu m}$ is
the length and $w\approx 1\,{\rm \mu m}$ the width of the wire. With
these numbers we obtain $L_k \approx 1.6\cdot 10^{-3}\,{\rm nH}$, which is
about three orders of magnitude smaller than (the linear part of) the
Josephson inductance. A simple estimate shows that this would lead to
a negative shift of the qubit resonance of only a few hundreds of
kHz. These corrections are negligible at the current level of
accuracy but can be easily included in the numerical simulation if necessary.

The aluminum 
antenna is evaporated on top of a sapphire substrate, the thickness of which is
$430\,{\rm \mu m}$ for samples $1$, $2$, $4$ and $5$ and $500\,{\rm \mu m}$
for samples $3$ and $6$. The contraction of aluminum with decreasing
temperature leading to a shrinkage of the cavity of about $0.5\%$ and
the reduction of the permittivity of sapphire by less than a per cent are taken into account~\cite{Pobel-1937,Krupka-1999}. 
\begin{figure}
\includegraphics[width=0.7\textwidth]{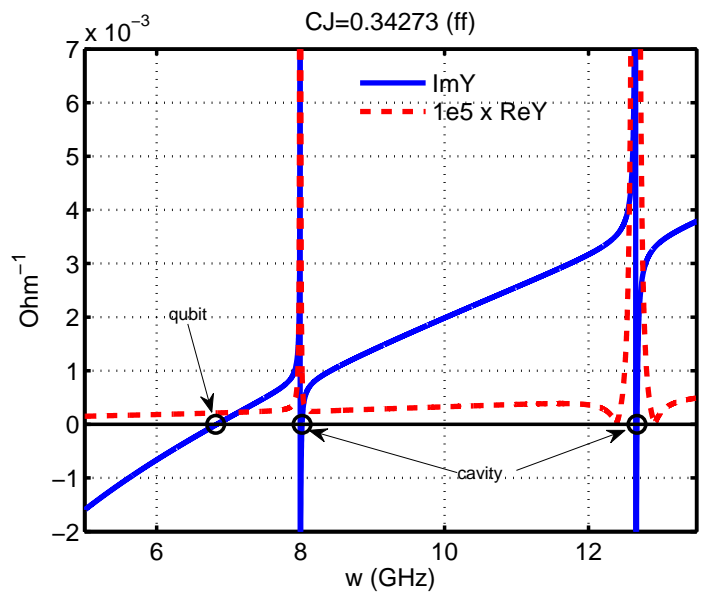}\caption{(Color online) Real and
  imaginary parts of the admittance $Y(\omega)=j\omega
  C_J-\frac{j}{\omega L_J}+Y_c(\omega)$ as obtained from the HFSS simulation.\label{fig:2b}}
\end{figure}

The imaginary and real parts of the resulting admittance $Y$ are shown in Fig.~\ref{fig:2b} for $C_J=0.34\,{\rm ff}$ over a range of
frequencies spanning three modes. 
The
lowest mode with the largest slope is identified with the qubit mode
and the remaining ones with cavity modes, although it must be kept in
mind that the states corresponding to these modes are all
superpositions of the bare modes. With this input, the corrections due to the
nonlinearity of the junction are computed as explained in the main
text. The results of the fit in $C_J$ are given in Table~I of the main
text and plotted in Fig.~\ref{fig:4b}. Details on the
measurement of the spectrum can be found in~\citet{Paik-2011a}.
\begin{figure}[ht]
\includegraphics[width=0.9\textwidth]{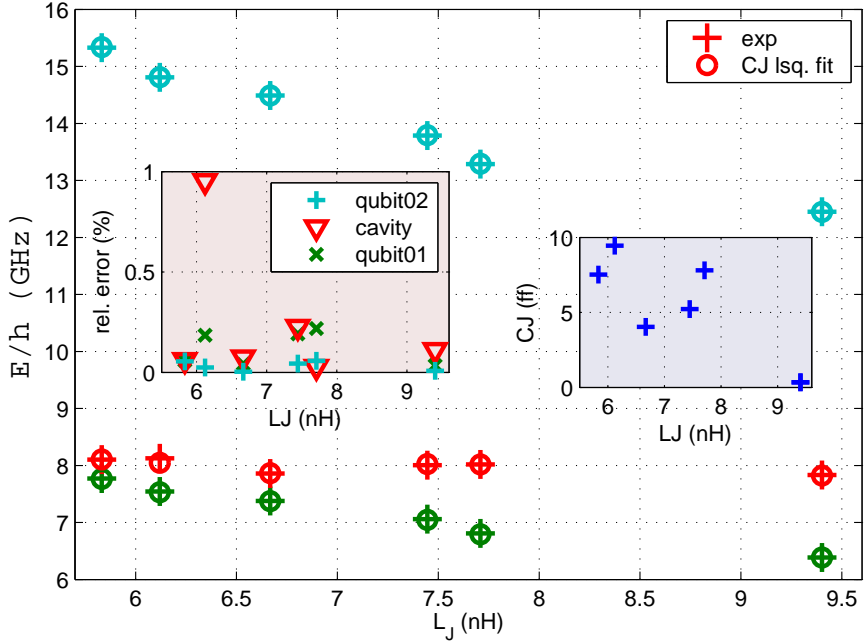}\caption{(Color
  online) Low-energy spectrum of six 3D-transmons. Theory values (open circles) are obtained by
fitting $C_J$ for each data set (stars). The green symbols correspond
to the $0\rightarrow 1$ qubit transition, the red symbols to the
lowest cavity resonance and the cyan symbols to the $0\rightarrow 2$
qubit transition. The left inset show the sub \% level
relative errors between theory and experiment and the right inset
shows the fitted values of $C_J$.\label{fig:4b}}
\end{figure}

\section{Black-box quantization with multiple Junctions}
For simplicity we focus on the dissipationless case. We consider a
system with $N$ Josephson junctions with {\em bare} Josephson energies $E_J^{(s)}$ and charging energies $E_C^{(s)}$, $s=1,\dots,N$, in parallel with a
common linear dissipationless but otherwise arbitrary electromagnetic
resonator as depicted in Fig.~\ref{fig:1b} (a). \begin{figure}[ht]
\begin{center}
\def\svgwidth{0.4\textwidth}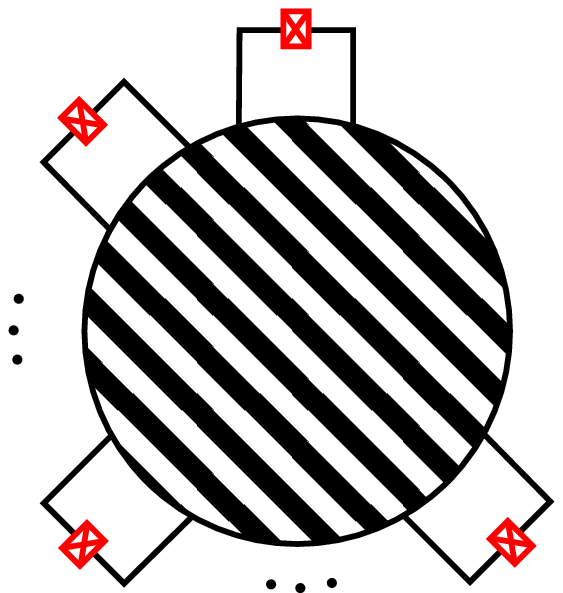\hspace{2cm}\def\svgwidth{0.4\textwidth}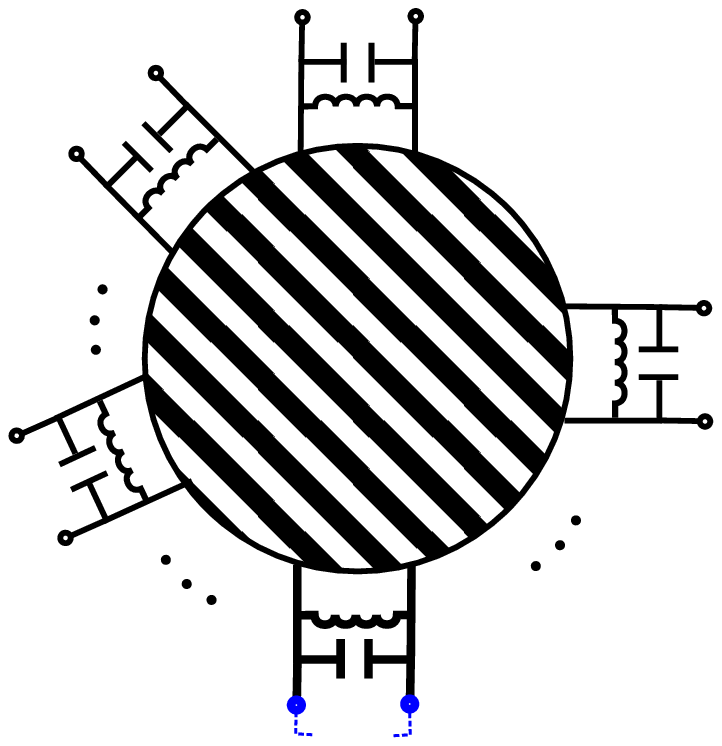\\\vspace*{0.8cm}
\def\svgwidth{0.9\columnwidth}
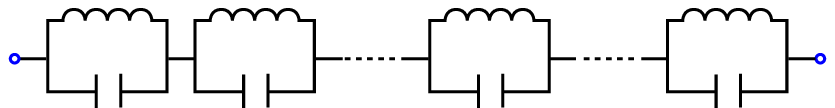
\caption{(Color
  online) (a) Schematics
of $N$ JJs (gray (red) boxed crosses) coupled to an
arbitrary linear circuit (striped disk). (b) Corresponding linearized
$N$-port circuit
with JJs replaced by parallel LC oscillators. (c)
Foster-equivalent circuit of the impedance $Z_{kk}(\omega)$ of the linearized circuit
shown in (b). The reference port $k\in\{1,\dots,N\}$ may be chosen arbitrarily.\label{fig:1b}}
\end{center}
\end{figure}
The unbiased isolated junctions alone are described
by the Hamiltonian
$H_J=\sum_{s=1}^N\left(4E_C^{(s)}(n_s)^2-E_J^{(s)}\cos(\varphi_s)\right)$, where
$n_s$ is the Cooper-pair number operator of the $s$-th junction
conjugate to the phase degree of freedom $\varphi_s$,
i.e. $[\varphi_s,n_s]=i\hbar$.\footnote{Note that strictly speaking the
commutator is rather $[\exp(i\varphi_s),n_s]=-\hbar\exp(i\varphi_s)$,
but as we neglect charge dispersion, it is consistent to neglect the
$2\pi$-periodicity of the commutation relation.} A
corresponding $N$-port {\em linear} circuit, shown in Fig.~\ref{fig:1b}~(b), is then defined by
associating a port with each junction and replacing the latter with a
parallel lumped element LC oscillator with inductance $L_J^{(s)} =
(\phi_0)^2/E_J^{(s)}$ and capacitance
$C_J^{(s)}=e^2/(2E_C^{(s)})$. Here and in the following
$\phi_0=\hbar/(2e)$ is the reduced flux quantum. This corresponds to
expanding the cosines in $H_J$ to second order in $\varphi_s$. We next
consider this linearized circuit classically.

A quantity of central importance in the following is the $N$-port
impedance matrix $\mathbf{Z}$ with elements $Z_{ss'}(\omega) =
V_s(\omega)/I_{s'}(\omega)\Big|_{I_i=0,i\not=s'}$. Let us choose
arbitrarily one reference port $k$ among the $N$ ports. By virtue
of Foster's theorem~\cite{Foster-1924} 
$Z_{kk}(\omega)$ is a purely
imaginary meromorphic function and can be synthesized by the
equivalent circuit of parallel LC oscillators in series shown in
Fig.~\ref{fig:1b}~(c). Explicitly
\begin{equation}\label{eq:4b}
Z_{kk}(\omega) = \sum_{p=1}^M\left(j\omega
    C_p^{(k)}+\frac{1}{j\omega L_p^{(k)}}\right)^{-1}\,,
\end{equation}
where $M$ is the number of modes and we have adopted the electrical engineering
convention of writing the imaginary unit as $j=-i$. This equivalent circuit mapping corresponds, in electrical
engineering language, to diagonalizing the
linearized system of coupled harmonic oscillators. Accordingly, the eigen-frequencies
$\omega_p=(L_p^{(k)}C_p^{(k)})^{-\frac{1}{2}}$ are determined by the
poles of $Z_{kk}$ or more conveniently by the
real roots of the admittance defined as $Y_k=Z_{kk}^{-1}$ and the effective capacitances are
determined by the frequency derivative on resonance of the latter as $C_p^{(k)} =(1/2)
{\rm Im}Y_k'(\omega_p)$. Note that~\cite{Foster-1924} ${\rm Im}Y_k'(\omega)
>0$. The Lagrangian of the system can be written as
\begin{equation*}
\mathcal{L}=\frac{1}{2}\sum_{p=1}^M\left(C_p^{(k)}(\dot\phi_p^{(k)}(t))^2+\frac{(\phi_p^{(k)}(t))^2}{L_{p}^{(k)}}\right)\,,
\end{equation*}
in terms of the normal (flux) coordinates
$\phi_p^{(k)}(t)=f_p^{k}e^{j\omega_p
  t}+(f_p^{k})^*e^{-j\omega_pt}$, associated with each of the
equivalent LC oscillators. From
this, we can
immediately write the Hamiltonian function of
the equivalent circuit as
$\mathcal{H}_0=2\sum_{p=1}^M(f_p^{k})^*(L_p^{(k)})^{-1}f_p^{k}$,
where the
subscript $0$ indicates that we consider the linear circuit
(Fig.~\ref{fig:1b} (b)). Note that the eigen-frequencies do not depend on
the choice of port, while the eigenmodes do. Kirchhoff's voltage law
implies that up to an
arbitrary constant,
$\varphi_k(t)=\phi_0^{-1}\sum_{p=1}^M\phi_p^{(k)}(t)$, where according to
Josephson's second relation,
$\varphi_k(t)=\phi_0^{-1}\int_{-\infty}^tV_k(\tau)d\tau$ is the phase
variable of the $k$-th (reference)
junction with voltage $V_k$. Importantly this simple relation
holds only for the junction at the reference port $k$. In order to
find the corresponding expressions for the other junctions ($s\not=k$), we notice that the AC voltage amplitude
$V_s(\omega)=j\omega\phi_s^{(k)}(\omega)$ at frequency
$\omega$ generated across port
$s$ in response to a current with amplitude $I_{s'}(\omega)$ applied at
port $s'$ is given by $V_s(\omega) = Z_{ss'}(\omega)I_{s'}(\omega)$. Hence we have
$\varphi_s^{(k)}(\omega)=(Z_{sk}(\omega)/Z_{kk}(\omega))\varphi_k(\omega)$. Combining
this with the above we find that
\begin{equation}\label{eq:5b}
\varphi_s^{(k)}(t)=\phi_0^{-1}\sum_{p=1}^M\frac{Z_{sk}(\omega_p)}{Z_{kk}(\omega_p)}\left(f_p^ke^{j\omega_pt}+(f_p^k)^*e^{-j\omega_pt}\right)\,.
\end{equation}
Quantization is achieve in the canonical way~\cite{Devoret-1995a,Clerk-2010a} by replacing the
flux amplitudes of the equivalent oscillators by operators as
\begin{equation}\label{eq:6b}
  f_p^{k(*)}\rightarrow\sqrt{\frac{\hbar}{2}\mathcal{Z}_{kp}^{\rm
      eff}}\,a_p^{(\dagger)}\,,\quad \mathcal{Z}_{kp}^{\rm
    eff}
  =\frac{2}{\omega_p{\rm
        Im}Y_k'(\omega_p)}\,,
\end{equation}
with the dimensionless bosonic annihilation (creation) operators
$a_p$ ($a_p^{\dagger}$). Direct substitution yields the Hamiltonian
$H_0=\sum_{l}\hbar\omega_la_l^{\dagger}a_l$ of $M$ uncoupled harmonic
oscillators (omitting the zero point energies) and the Schr{\"o}dinger
operator of phase of the $l$-th junction is
\begin{equation}\label{eq:7b}
\hat\varphi_s^{(k)} =\phi_0^{-1}
\sum_{p=1}^M\frac{Z_{sk}(\omega_p)}{Z_{kk}(\omega_p)}\sqrt{\frac{\hbar}{2}\mathcal{Z}_{kp}^{{\rm
      eff}}}\left(a_p+a_p^{\dagger}\right)\,.
\end{equation}
This is Eq.~(7) of the main text using that
$\hat\phi_s^{(k)}=\phi_0\hat\varphi_s$. The superscript makes explicit
the
dependence on the reference port. Accordingly the root mean
square fluctuation of the flux of junction $s$ in the multi-mode Fock state
$\ket{n_1,n_2,\dots,n_M}$ is given by $\sqrt{\braket{(\hat\phi_s^{(k)})^2}}
= \frac{\hbar}{2}\sum_{p=1}^M\left(\frac{Z_{sk}(\omega_p)}{Z_{kk}(\omega_p)}\right)^2\mathcal{Z}_{kp}^{{\rm
      eff}}\left(1+2n_p\right)$.

The anharmonic terms generated by the non-linearity of the Josephson
inductance, necessary to build a qubit, are included by expressing the
higher order terms in
the expansion of the cosine in the harmonic basis. Including up to the quartic
terms we obtain explicitly after normal ordering
\begin{align}\label{eq:1b}
H&=H_0-\sum_{pp'}\gamma_{pp'}\left(2a_p^{\dagger}a_{p'}+a_p^{\dagger}a_{p'}^{\dagger}+a_pa_{p'}\right)\\
&\quad-\sum_{pp'qq'}\beta_{pp'qq'}\left(6a_p^{\dagger}a_{p'}^{\dagger}a_qa_{q'}+4a_p^{\dagger}a_{p'}^{\dagger}a_q^{\dagger}a_{q'}+4a_p^{\dagger}a_{p'}a_qa_{q'}+a_pa_{p'}a_qa_{q'}+a_p^{\dagger}a_{p'}^{\dagger}a_q^{\dagger}a_{q'}^{\dagger}\right)+\sum_{s=1}^N\mathcal{O}({\hat\varphi_{s}}^6)\nonumber\,,
\end{align}
with coefficients
$\beta_{pp'qq'}=\sum_{s=1}^N\frac{e^2}{24L_J^{(s)}}\xi_{sp}\xi_{sp'}\xi_{sq}\xi_{sq'}$
and $\gamma_{pp'}=6\sum_{q=1}^M\beta_{qqpp'}$ where, choosing the first port as the reference port, $\xi_{sp}=\frac{Z_{s1}(\omega_p)}{Z_{11}(\omega_p)}\sqrt{\mathcal{Z}_{1p}^{{\rm
      eff}}}$. Treating the $\varphi^4$ nonlinearity in first order
perturbation theory, one obtains the expressions for the energy,
generalized chi-shift and generalized anharmonicity
given by Eq.~(\ref{eq:3})) of the main text.
\end{widetext}

\end{document}